\documentclass[a4paper]{JHEP3}
\usepackage[centertags]{amsmath}
\usepackage{amssymb}
\def\bs{{\bf s}}
\def\bv{{\bf v}}

\def\bCV{{\mathfrak V}}
\def\bCS{{\mathfrak S}}

\def\Ss#1{\bs{\mathnormal #1}}
\def\ST#1{\bCS{\mathnormal #1}}
\def\V#1{\bv{\mathnormal #1}}
\def\VT#1{\bCV{\mathnormal #1}}

\def\del{\nabla}
\newcommand{\beq}{\begin{equation}}
\newcommand{\eeq}{\end{equation}}
\newcommand{\beqa}{\begin{eqnarray}}
\newcommand{\eeqa}{\end{eqnarray}}
\newcommand{\be}{\begin{equation}}
\newcommand{\ee}{\end{equation}}
\newcommand{\bea}{\begin{eqnarray}}
\newcommand{\eea}{\end{eqnarray}}
\newcommand{\bA}{\begin{array}}
\newcommand{\eA}{\end{array}}
\newcommand{\bc}{\begin{center}}
\newcommand{\ec}{\end{center}}

\title{Forced Fluid Dynamics from Gravity}

\author{{Sayantani Bhattacharyya, R. Loganayagam, Shiraz Minwalla,
 Suresh Nampuri,\linebreak
 Sandip P.  Trivedi }\\ 
{Dept. of Theoretical Physics, Tata Institute of Fundamental Research,\linebreak
Homi Bhabha Rd, Mumbai 400005, India.}\\
E-mail: \email{sayanta, nayagam, minwalla, suresh, sandip @theory.tifr.res.in}}

\author{Spenta R. Wadia \\ 
{International Centre for Theoretical Sciences and \linebreak
Dept. of Theoretical Physics, Tata Institute of Fundamental Research,\linebreak
Homi Bhabha Rd, Mumbai 400005, India.}\\
E-mail: \email{wadia@theory.tifr.res.in}}

\abstract{We generalize the computations of \cite{Bhattacharyya:2008jc} to 
generate long wavelength, asymptotically 
locally AdS$_5$ solutions to the Einstein-dilaton 
system with a  slowly varying boundary dilaton field and a weakly curved boundary metric. 
Upon demanding regularity , our solutions are dual, under the AdS/CFT correspondence, to arbitrary fluid flows in the boundary theory formulated on a weakly curved manifold 
with a prescribed slowly varying coupling constant. These solutions turn out to be parameterized by four-velocity and temperature fields that are constrained to obey the boundary covariant Navier Stokes equations with a dilaton dependent forcing term. We explicitly evaluate the stress tensor and Lagrangian as a function of the velocity, 
temperature, coupling constant and curvature fields, to second order in the 
derivative expansion and demonstrate the Weyl covariance of these expressions. We also construct the event horizon of the dual solutions to second order in the derivative expansion, and use the area form on this event horizon to construct an entropy current for 
the dual fluid. As a check of our constructions we expand the exactly known 
solutions for rotating black holes in global AdS$_5$ in a boundary derivative 
expansion and find perfect agreement with all our results upto second 
order. We also find other simple solutions of the forced fluid 
mechanics equations and discuss their bulk interpretation. Our results may 
aid in determining a bulk dual to  forced flows exhibiting  steady state turbulence.}

\preprint{TIFR/TH/08-25}

\begin{document}



\section{Introduction}

The gauge gravity correspondence\cite{Maldacena:1997re} is one of the key ideas to have emerged from developments in string theory. It has already led to considerable progress, 
and promises to teach us much more both about quantum gravity and about 
strongly coupled field theory dynamics.
 
The best studied class of gauge gravity dualities relate the dynamics 
of a particular conformal field theory on $R^{3,1}$ to the 
physics of a corresponding gravitational theory on an AdS$_5$ background. We now have an infinite number of proposed dualities of this nature. As quantum field theories may be formulated on arbitrary base manifolds it is natural to attempt to generalize these
$AdS/CFT$ correspondences to obtain dual descriptions of these CFTs on an arbitrary Lorentzian base manifold $M_{3,1}$.  

According to the usual rules of the $AdS/CFT$ correspondence (See \cite{Aharony:1999ti} for a review), the classical phase space of the dual bulk description is the set of regular 
solutions of the relevant bulk equations of motion that asymptote at small 
$z$ (the boundary) to the metric\footnote{Together with similarly prescribed 
asymptotic conditions for all other fields.} 
\begin{equation}\label{asympmetric}
ds_5^2=\frac{dz^2+ ds_{3,1}^2}{z^2}
\end{equation}
where $ds_{3,1}^2$ is the metric of $M_{3,1}$(See Appendix \ref{app:notation} for a list of notation employed in this paper). In this paper we use the methods presented in 
\cite{Bhattacharyya:2008jc,Bhattacharyya:2008xc} (see also  \cite{VanRaamsdonk:2008fp,Dutta:2008gf}) to completely characterize a special corner of this phase space that is dual, under the 
$AdS/CFT$ correspondence,  to boundary fluid dynamics.

Through most of this paper, we focus on the bulk dual of the four-dimensional $CFT$ dynamics at large $N$ and infinitely strong coupling formulated on a weakly curved four dimensional spacetime \footnote{In this introduction, we will first describe our results assuming no variation in the boundary coupling constant, but we will later generalize these results to a prescribed weakly varying coupling constant.}. 
Away from possible singularities, the the bulk equations 
that determine dual dynamics in this limit may be derived from a two derivative action that describes the interaction of gravity with other fields \footnote{For example, when the
CFT in question is ${\cal N}=4$ Yang Mills theory, the action corresponds to
the $S^5$ compactification of IIB supergravity.} The equations of motion of all such systems have a universal subsector, in which the Einstein frame metric is a solution of Einstein gravity with a negative cosmological constant and all other fields are simply zero\footnote{A class of two-derivative gravitational theories also admit a consistent truncation to a larger Einstein-dilaton system which we will employ to describe later the bulk duals of boundary theories with prescribed variation of the coupling constant}. In this paper, we study this universal subsector subject to the boundary conditions \eqref{asympmetric}. 

We are specifically interested in regular bulk configurations  whose variations in the boundary directions are slow compared to the length scale set by the local energy density of the solution. Experience with field theory along with AdS/CFT suggests that dynamics of such long-wavelength solutions is  effectively described by four-dimensional hydrodynamics.\footnote{In general, the fluid mechanics approximation is self consistent only when the length scale associated with curvatures of the boundary metric $ds_{3,1}^2$,  and the length scale of variation of the boundary coupling constant,  are also large in units of the same energy density scale. We assume that these  conditions are met  in the following analysis.}. Starting from the bulk Einstein equations, in this paper we demonstrate that this is indeed the case. More specifically, within a boundary derivative expansion, we construct an explicit map from the solutions of Navier Stokes equations (with distinguished, gravitationally determined values of parameters) on the manifold $M_{3,1}$ 
to the space of regular solutions of Einstein's equations with a negative 
cosmological constant that asymptote to \eqref{asympmetric} at small $z$. \footnote{See \cite{Policastro:2001yc,
Son:2002sd,Policastro:2002se,Herzog:2002fn,Policastro:2002tn,Herzog:2002pc,
Herzog:2003ke,Kovtun:2003vj,Buchel:2003ah,Kovtun:2003wp,Buchel:2003tz,
Kovtun:2004de,Buchel:2004hw,Buchel:2004di,Buchel:2004qq,
Kovtun:2005ev,Starinets:2005cy,Aharony:2005bm,Janik:2005zt,
Mas:2006dy,Son:2006em,Saremi:2006ep,Maeda:2006by,CasalderreySolana:2006rq,
Gubser:2006bz,Janik:2006gp,Liu:2006nn,Nakamura:2006ih,Janik:2006ft,
Benincasa:2006fu,Lin:2006rf,Saremi:2007dn,Heller:2007qt,Son:2007vk,
Kovchegov:2007pq,Lahiri:2007ae,Myers:2007we,Bhattacharyya:2007vs,
Lin:2007fa,Chesler:2007sv,Ejaz:2007hg,Kats:2007mq,Benincasa:2007tp,
Baier:2007ix,Bhattacharyya:2008jc,Natsuume:2007ty,Natsuume:2007tz,
Kajantie:2008rx,Natsuume:2008iy,Loganayagam:2008is,Buchel:2008ac,
VanRaamsdonk:2008fp,Choi:2008he,Gubser:2008vz,Bhattacharyya:2008xc,Chernicoff:2008sa,
Buchel:2008xr,Myers:2008fv,Dutta:2008gf,Siopsis:2008xz,Gubser:2008pc,Heller:2008mb} for a list of relevant literature on this subject.}

The method employed in this paper is a direct generalization of the 
procedure employed in \cite{Bhattacharyya:2008jc, Bhattacharyya:2008xc}
(see also \cite{VanRaamsdonk:2008fp}). Given a velocity field $u^\mu(x)$ and a temperature field $T(x)$ we note that the metric 
\begin{equation}\label{genbbn1} \begin{split}
ds^2& = -2 u_\mu(x^\mu) dx^\mu dr - r^2 f(b(x^\mu)r) u_\mu u_\nu dx^\mu dx^\nu 
+ r^2 {\cal P}_{\mu\nu} dx^\mu dx^\nu \\
{\cal P}_{\mu\nu}&=g_{\mu\nu}(x^\mu)+u_\mu(x^\mu) u_\nu(x^\mu) \\
f(r)&=1-\frac{1}{r^4}\\
\end{split}
\end{equation}
(where $g_{\mu\nu}(x^\mu)$ is the metric on $M_{3,1}$) has several desirable properties. First, its asymptotic form  matches \eqref{asympmetric}. Next note that lines of constant $x^\mu$ in \eqref{genbbn1} are ingoing null geodesics. Tubes centered around these geodesics are locally similar to tubes in a uniform black-brane geometry but with a temperature and velocity that varies with $x^\mu$. In fact, this metric reduces to the exact uniform black-brane solution of Einstein's equations when $u^\mu(x)$ and $T(x)$ and $g_{\mu\nu}(x)$ are constants. 
Finally, under suitable assumptions regarding the late-time behavior of the boundary-metric, boundary velocity,
 and the boundary temperature fields,  this metric  has a regular future event-horizon.  

We then generalize the procedure of \cite{Bhattacharyya:2008jc} to demonstrate that \eqref{genbbn1} may be thought of as the first term in a systematic expansion of Einstein equations in a 
power series in $\frac{1}{TL}$ where $T$ is the local temperature of the 
solution and $L$ the local length scale of variation of the temperature and 
velocity and metric functions in the boundary directions.

We work out the rules of this perturbation theory to arbitrary order in $\epsilon$, and 
explicitly implement these rules to second order. As in \cite{Bhattacharyya:2008jc}, this 
procedure works only when the velocity and temperature fields obey a particular dynamical equation of motion. This equation of motion turns out simply to be $\del_\mu T^{\mu\nu}=0$ (the indices ($\mu, \nu$) run only over the boundary directions ) where the boundary stress tensor $T^{\mu\nu}$ is 
a given function of the velocity and temperature fields, whose form is determined, order by 
order in the derivative expansion, by the perturbative procedure itself. Of course this equation of motion is simply the Navier Stokes equation of fluid dynamics, with particular values of fluid parameters predicted by gravity. Our explicit second order solution to this perturbation theory 
yields an explicit expression for the fluid dynamical stress tensor to second order in the derivative expansion. Most terms in this stress tensor are obtained by the simple covariantization of the 
flat space second order stress tensor reported in \cite{Bhattacharyya:2008jc}. However in addition 
we find a contribution proportional to a curvature tensor that the analysis of \cite{Bhattacharyya:2008jc} was blind to. The existence of such a term was already predicted in \cite{Baier:2007ix}, and our result is in complete agreement with their 
prediction (including coefficients) providing a check both of our results
as well as those of \cite{Baier:2007ix}.

We show on general grounds that the metrics constructed by the perturbative procedure developed in this paper must transform in a specified fashion under boundary Weyl transformations. In particular, Weyl covariance demands that the metric dual to a fluid configuration be of the form
\begin{equation}\label{weylmet:eq}
\begin{split}
ds^2 &= -2 W_1 \ u_\mu dx^\mu(dr+r \mathcal{A}_\lambda dx^\lambda) \\
&+\left[ r^2\left( W_2 \ g_{\mu\nu} + W_3 u_\mu u_\nu  \right) + r\left( W_{4\mu} u_\nu + u_\mu W_{4\nu}\right) + W_{5\mu\nu}  \right] dx^\mu dx^\nu . \\
\mathcal{A}_\lambda&\equiv u.\nabla u_\lambda -\frac{\nabla.u}{3} u_\lambda
\end{split}
\end{equation}
where $W_1,W_2$ and $W_3$ are Weyl-invariant scalars, $W_{4\mu}$ is a Weyl-invariant transverse vector and $W_{5\mu\nu}$ is a Weyl-invariant transverse traceless tensor and $\mathcal{A}_\lambda$ is the fluid mechanical Weyl connection introduced in \cite{Loganayagam:2008is}. This requirement, while logically necessary is not algebraically automatic in our perturbative procedure, and so gives a second nontrivial check on our results.  Our construction passes this test;  in fact we are able,in $\S$\ref{sec:weyl}, to rewrite the bulk metric dual to second order fluid dynamics  entirely in terms of the Weyl covariant formalism of \cite{Loganayagam:2008is}, making the Weyl transformation properties of our solution manifest. 

Having constructed these bulk metrics dual to fluid dynamics, we then proceed to study their causal properties. In particular, we generalize the results of \cite{Bhattacharyya:2008xc}  to  demonstrate that (under appropriate assumptions) all ($r=0$) singularities present in the solutions presented in this paper are shielded from the boundary of $AdS$ by a regular event horizon. In fact, we find a local expression for the radial location of the event horizon as a function of boundary fluid data, generalizing the results of \cite{Bhattacharyya:2008xc}. We then use our construction of the event horizon together with the pull back of the area form on the horizon to the boundary (see \cite{Bhattacharyya:2008xc}) to construct an entropy current for the curved space fluid dynamics constructed in this paper. The non-negativity of the divergence of our entropy current is guaranteed by the area increase theorem of general relativity.

As an additional check and application of our results we go on to study the most general AdS$_5$ Kerr solutions \cite{Hawking:1998kw,Awad:1999xx,Awad:2000aj,Gibbons:2004uw,Gibbons:2004ai,Chong:2005hr,Cvetic:2005zi} in fluid dynamical terms. These exact solutions are labelled by their mass and two angular velocities. It was already observed in \cite{Bhattacharyya:2007vs} that the stress tensor and thermodynamical properties of these black holes agree with the expectations of perfect fluid dynamics in the large mass limit. In this paper we go a step further; we rewrite the bulk  metric of these black holes in the coordinates  employed in our general construction of bulk duals to fluid flows. In these co-ordinates the AdS$_5$ Kerr solution assumes a particularly simple (and a manifestly Weyl-covariant) form 
\begin{equation} \label{bhfrintro:eq}
\begin{split}
ds^2 &= -2 u_\mu dx^\mu (dr+r \mathcal{A}_\lambda dx^\lambda) + r^2 g_{\mu\nu} dx^\mu dx^\nu\\
&-\left(u_\mu \mathcal{D}_\lambda\omega^\lambda{}_\nu + \omega_\mu{}^\lambda\omega_{\lambda\nu} + \frac{\mathcal{R}}{6} u_\mu u_\nu \right)dx^\mu dx^\nu + \frac{2 m}{r^2}\left(1+\frac{1}{2r^2} \omega_{\alpha\beta}\omega^{\alpha\beta}\right)^{-1} u_\mu u_\nu dx^\mu dx^\nu \\
\end{split}
\end{equation}
where $\omega_{\mu\nu}$ is the vorticity of the fluid, $\mathcal{D}$ is the fluid mechanical Weyl covariant derivative (associated with the Weyl connection $\mathcal{A}_\lambda$) and $\mathcal{R}$ is the Weyl-covariantized Ricci scalar. In this form, this bulk metric admits a simple and explicit all orders expansion in fluid dynamical terms. The expansion of \eqref{bhfrintro:eq} to second order 
in the derivative expansion is exactly reproduced  by our general metric dual to fluid dynamics, upon substitution of the velocity field $u_\mu$ into our metric\footnote{It is interesting to note that in the special case of solutions with vorticity $\omega_{\mu\nu}= 0$, our metric exactly reproduces the AdS Schwarzschild blackhole.}. We also verify that our formulas for the location of the event horizon and the local entropy current of this black hole match the exact formulas for the same quantities upto second order in the derivative expansion. 

As we have explained above, we have constructed gravitational solutions dual to every solution of the covariant Navier Stokes equations 
\begin{equation}\label{nseq}
\nabla_\mu T^{\mu\nu}=0
\end{equation} 
on an arbitrary background manifold with metric $g_{\mu\nu}$. Let us now
consider a special case of this equation; let 
\begin{equation}  \label{metricvar}
g_{\mu\nu}=(g_0)_{\mu\alpha}\left( \delta^{\alpha}_\nu +h^{\alpha}_\nu \right)
\end{equation}
 for small $h^\alpha_\nu$ and work to 
first order in $h^\alpha_\nu$. The Navier Stokes equation may be rewritten as 
\begin{equation}\label{forcens}
(\nabla_0)_\mu (T_0)^{\mu\nu}= f^\nu
\end{equation}
where $(T_0)^{\mu\nu}$ is the stress tensor of the corresponding fluid flow 
in the  space-time with metric $(g_0)_{\mu\nu}$, and the 
effective forcing function $f^\nu$ is given by 
\[f_\nu =-\frac{1}{2} (T_0)^\alpha_\nu \partial_\alpha h - \partial_\mu [(T_0)^{\mu\alpha}h_{\alpha\nu}]+\frac{1}{2} (T_0)^{\alpha\beta}
 \partial_\nu h_{\alpha\beta} - \nabla_\mu^{(0)}[\delta T^{\mu\lambda}g^{(0)}_{\lambda\nu}]
\]
All indices in this equation are now raised and lowered with the metric 
$(g_0)_{\mu\nu}$ and $\nabla_0$ is the covariant derivative with respect to the 
same metric. In particular, for a symmetric tensor $Q^{\mu\nu}$ we have
\begin{equation}
\nabla^{(0)}_\mu Q^{\mu}{}_\nu \equiv \frac{1}{\sqrt{-g^{(0)}}} \partial_\mu \left[ \sqrt{-g^{(0)}}\ Q^{\mu}{}_\nu\right] -\frac{1}{2} Q^{\alpha\beta}\partial_\nu g^{(0)}_{\alpha\beta}
\end{equation}
In the expression above, $\delta T^{\alpha\nu}$  is the first order change 
in the fluid stress tensor  \eqref{stetcn}, under the variation of the metric given in 
\eqref{metricvar}, and the variation in the four-velocity $\delta u^\mu= -\frac{u^\mu}{2} 
u^\alpha u^\beta h_{\alpha\beta}$, keeping the temperature fixed. For instance,   $\delta T^{\alpha\nu}$ for the perfect fluid stress tensor $T^{\alpha\nu}=T^4(4 u^\alpha u^\nu +g^{\alpha\nu})$ is given by 
\begin{equation}
\delta T^{\mu\nu} = - T^4\left[g^{(0)\mu\alpha}h_{\alpha\beta}g^{(0)\beta\nu}
+ 4 u^\mu u^\nu h_{\alpha\beta} u^\alpha u^\beta \right]
\end{equation}

By choosing $h_{\mu\nu}$ appropriately we can produce a wide range of forcing functions $f_\mu$ with one qualitative restriction; $f^\mu$ is relatively mild as it is necessarily at least of first 
order in the boundary derivative expansion. Nonetheless we expect even 
this mild forcing function to be able to  stir the fluid into flows with 
velocity differences $v$ of unit order. The reason for this expectation is 
that that every term on the LHS of \eqref{forcens} is also of at least of 
first order in derivatives, so that factors of $1/L$ should `cancel out' 
between the LHS and RHS. In $\S$\ref{sec:simp}, below we present evidence for this 
assertion by presenting some simple steady state solutions to the equations 
of fluid dynamics that are pushed into flows with large velocities. 

The Reynolds number for flows of the conformal fluids studied in this paper 
is given by $T L v$; where $L$ is the length scale of variation of velocities as above. 
If $v$ can indeed be stirred to order unity, it follows that the Reynolds numbers of the corresponding flows are very large in the limit $TL \gg 1$ considered in this paper. Recall that fluid flows with 
large Reynolds numbers are expected to be turbulent. The discussion of this paragraph suggests that it should be possible to choose $h_{\mu\nu}$ and correspondingly the functions $f^\nu$ to stir the boundary fluid into, for instance, steady state turbulence. The map from fluid dynamics to gravity, presented in this paper then yields a bulk dual to this turbulent fluid flow. It would be very interesting to explore this in more detail, and in particular to investigate whether bulk dual considerations could shed 
new light on the apparent universality of turbulent flows. 

In this introduction so far, we have described the  construction of the bulk 
solution dual to the long wavelength dynamics of a field theory on an arbitrary 
weakly curved manifold. This discussion can be generalized to the bulk solution  dual to  
 an arbitrary fluid dynamical flow of the field theory, on an arbitrary weakly curved manifold, and with an 
arbitrary slowly varying coupling constant. In order to do this we focus on the class of gravitational theories which admit consistent truncation to Einstein-dilaton sector; this includes all supergravity 
theories that result from the compactification of IIB supergravity down to AdS$_5$, and so includes the duals to all well understood examples of the AdS$_5$/CFT$_4$ correspondence. 

For this class of theories, we determine the dual to the the fluid dynamical motions of field theories with actions of the form  
\begin{equation}\label{newact}
S=\int \sqrt{g} e^{-\phi}\ {\cal L}
\end{equation}
where $\phi(x^\mu)$ is an arbitrarily specified slowly varying function.
\footnote{In sections $\S$\ref{sec:deriv:exp} and $\S$\ref{sec:exact} below, we economize on space by presenting only our most general results which account for the effects of a varying dilaton field. The bulk metric dual to fluid flows on a curved spacetime with constant dilaton may easily be obtained 
from the results of $\S$\ref{sec:deriv:exp} and $\S$\ref{sec:exact} by simply setting the boundary dilaton to a constant in all equations.}
As the Lagrangian \eqref{newact} explicitly breaks translational invariance, 
the equations of motion obeyed by the velocity and temperature fields of 
this system are modified; in fact the Noether procedure for translational 
invariance, applied to \eqref{newact}, yields the equation 
\begin{equation}\label{noncons} \begin{split}
\nabla_\mu T^{\mu\nu}&=  e^{-\phi}\ {\cal L}\ \nabla^\nu \phi\\
T_{\mu\nu}& = -\frac{2}{\sqrt{g}} \frac{\delta}{ \delta g^{\mu\nu}} S
\end{split}
\end{equation}
i.e. the Navier Stokes equations with an additional explicit forcing term. This equation also 
follows from a direct analysis of the Einstein-dilaton system near the boundary, upon 
employing the usual holographic formulas for $T_{\mu\nu}$ and the Lagrangian (see Appendix 
\ref{app:forcing}).
 
In this paper, we perturbatively determine gravity dual descriptions of 
fluid flows that obey \eqref{noncons} and in the process determine 
expressions for the stress tensor and the Lagrangian, as a function of the 
velocity fields, temperature field, curvature and coupling constant, order 
by order in the derivative expansion. The procedure we employ to derive these results is a straightforward generalization of the procedure described above for arbitrary weakly curved boundary metric. We search for solutions of the Einstein-dilaton system with boundary conditions on the metric field as described above, but additionally require that the dilaton field $\Phi(x^\mu)$ asymptote to $\phi(x^\mu)$. The configuration \eqref{genbbn1}
supplemented with $\Phi(x^\mu)=\phi(x^\mu)$ once again turns out to be a suitable first term in a systematic perturbative expansion of solutions of the Einstein-dilaton system in powers of $\epsilon$.
This perturbation theory works only if the integrability condition \eqref{noncons} is obeyed. When this equation is met, the perturbative procedure generates expressions for $T^{\mu\nu}$ and ${\cal L}$ as functions of velocities, temperatures and background fields. We have explicitly worked out this perturbation theory - and consequently the corresponding expressions for these quantities - explicitly to second order in derivatives. Under suitable assumptions regarding the late-time behavior of the boundary-metric and the variation of the coupling constant, we have also determined the location of the event horizon and an expression for the entropy current for this larger class of bulk configurations.

Upon solving the gravitational equations it turns out that the expectation value of the Lagrangian is itself proportional to a derivative of the dilaton; as a consequence, the explicit forcing function in \eqref{noncons} is of second order or higher in boundary derivatives, and so is milder than the effective forcing function described by a metric fluctuation \eqref{forcens}. Nonetheless 
fluid flows forced by the dilaton are interesting for their bulk interpretation. 

While we have explicitly derived a duality between solution in fluid mechanics of strongly coupled gauge CFTs and solutions of the equations of gravity, we expect this duality to continue to provide a map between fluid flows in the CFT at arbitrary nonzero coupling and solutions to the equations of the appropriate dual classical string theory. The equations of fluid dynamics presented in this paper have been computed from Einstein gravity and so are only valid at large 'tHooft coupling,  $\lambda\equiv e^{\phi}N$. However, many of the conclusions in this paper only depend on the fact that the forcing function has a mild effect, if the external gravitational and dilaton fields vary slowly enough compared to the temperature. This feature is determined by symmetry considerations and must be valid 
at arbitrary $\lambda$. 

With this in mind, consider the evolution of the hot boundary field theory with a time dependent
dilaton that starts at a large value, dips at a particular time to a value
of order unity, and then climbs back to a large value at later times. It
seems reasonable to guess that the equations of fluid dynamics derived in this paper
provide qualitative (though not quantitative) guidance to the nature of the evolution
of a hot classical solution of string theory through regions of string scale
curvature\footnote{Recall that the curvature in string units of the corresponding string
backgrounds is known to scale as $\lambda^{-\frac{1}{4}}$ at strong coupling
and is expected to be of unit order when $\lambda$ is of unit order. }. 
In section \S\ref{sec:simp} we discuss such an evolution which turns out to be rather
mild, suggesting the absence of qualitative surprises in at least this
particular excursion through string scale curvatures.


To conclude this introduction, we reemphasize that in this paper we have determined the 
gravitational dual description of the motion of a forced fluid. 
Forced fluid flows have been the subject of intense investigation for 
over a century\cite{1959flme.book.....L} and display fascinating dynamical behaviors 
all of which must be inherited by  gravity. We hope that this connection can be used
to extract useful lessons for the study of gravity and/or fluid dynamics.

\section{Long wavelength solutions of the 
Einstein-dilaton system in derivative expansion}\label{sec:deriv:exp}

Consider the Einstein-dilaton system with a negative cosmological constant and boundary conditions as described in the introduction. In this section, we will describe how one may systematically solve for a class of long wavelength solutions of this system order by order in $\epsilon=\frac{1}{TL}$, where 
$L$ is the length scale of variation of the dilaton field, boundary metric, velocity and temperature fields and $T$ is the local temperature of the fluid. All the results of this section apply to any CFT whose bulk dual description admits a consistent truncation to the equations of motion that follow from the Lagrangian
\begin{equation}\label{eindil}
S=\frac{1}{16 \pi G_5} \int \sqrt{G}\left( R +12 -\frac{1}{2} 
(\partial \Phi)^2 \right)
\end{equation} 
This property is true of all effective actions that are obtained from the 
compactification IIB theory on an Einstein manifold, and so of every one 
of the infinitely many known examples of the AdS/CFT correspondence. 

Notice that the Lagrangian \eqref{eindil} admits a further consistent 
truncation to Einstein gravity with a negative cosmological constant if 
we set the dilaton field to any constant value. Consequently, when our 
boundary conditions allow us to set the dilaton to a constant the equations 
from \eqref{eindil} further simplify. All the results of this section and 
the next apply with greater universality in this limit - they apply to any CFT whose
dual description is a two derivative theory of gravity interacting with other fields. 

\subsection{Uniform Branes in the Einstein-dilaton System}

The equations of motion that follow from Lagrangian \eqref{eindil}
are given by

\begin{equation}\label{eom}
\begin{split}
& R_{MN}-\frac{\partial_M \Phi \partial_N \Phi}{2} 
-\frac{G_{MN}}{2}\left( R + 12-\frac{(\partial \Phi)^2}{2} 
\right) = 0\\
&\implies R=\frac{(\partial \Phi)^2}{2} -20, \qquad
R_{MN}-\frac{\partial_M \Phi \partial_N \Phi}{2} + 4 G_{MN}=0\\
& \nabla^2 \Phi=0.\\
\end{split}
\end{equation}
 
There exists a well known four parameter class of exact solutions to these equations of 
motion 
\begin{equation}\label{genbb} \begin{split}
ds^2& = -2 u_\mu dx^\mu dr - r^2 f(br) u_\mu u_\nu dx^\mu dx^\nu 
+ r^2 {\cal P}_{\mu\nu} dx^\mu dx^\nu \\
\Phi&=\Phi_0 \\
{\cal P}_{\mu\nu}&=g_{\mu\nu}+u_\mu u_\nu \\
f(r)&=1-\frac{1}{r^4}\\
\end{split}
\end{equation}
where $g_{\mu\nu}$ is an arbitrary constant matrix of signature $(-1,1,1,1)$, 
$b$ is a constant and $u_\mu$ is a constant unit normalized velocity vector : 
$u_\mu u_\nu g^{\mu\nu}=-1$.

\subsubsection{Regulation and Weyl Frames}

In this subsection, we pause to recall a well known subtlety in the boundary 
interpretation (via the $AdS/CFT$ correspondence) of locally asymptotically $AdS$ metrics like \eqref{genbb}. This interpretative subtlety will play no role in the calculation we describe in this section or the next. However, it will permit a stringent test on the self-consistency of our final results in $\S$~\ref{sec:weyl}. 

We begin by noting that the metric \eqref{genbb} asymptotes to 
\begin{equation}\label{asymp} 
ds^2 = -2 u_\mu dx^\mu dr + r^2 g_{\mu\nu} dx^\mu dx^\nu 
\end{equation}
and according to the usual rules of the AdS/CFT correspondence, describes a state in the
CFT  formulated on a space with any of the infinite numbers of metrics  that are Weyl equivalent to $g_{\mu\nu}$. 
. 
Next, we note that in order to find the dual interpretation of any asymptotically $AdS$ space, it is convenient to regulate the solution near its boundary and the bulk solution may then be regarded as a state on the field theory on a base manifold whose metric is proportional to the induced metric on this regulated boundary. 

In particular, we may choose to regulate the boundary of \eqref{asymp} on slices of large constant $r$. With this choice the dual CFT resides on a space whose metric may be taken to be precisely $g_{\mu\nu}$, and \eqref{genbb} in fact represents the CFT on a space with metric 
$g_{\mu\nu}$ uniform temperature $T=\frac{1}{\pi b}$ and 
in uniform motion at velocity $u^\mu$.Alternatively, we may choose to regulate the boundary of \eqref{genbb} along slices of constant ${\tilde r}=e^{-\chi(x^\mu)} r$ for any arbitrary 
function $\chi(x^\mu)$. In order to interpret this new slicing it is convenient to rewrite the metric in terms of ${\tilde r}$; Asymptotically, we have
\begin{equation}\label{asympn1} 
ds^2 = -2 {\tilde u}_\mu dx^\mu d{\tilde r} + {\tilde r}^2 {\tilde g}_{\mu\nu} 
dx^\mu dx^\nu 
\end{equation}
where 
${\tilde u}_\mu=e^{\chi} u_\mu$ and ${\tilde g}_{\mu\nu}=e^{2 \chi}
{\tilde g}_{\mu\nu}$.  
It follows that the constant ${\tilde r}$ slicing of  \eqref{genbb} describes a state the CFT on with spacetime varying background metric ${\tilde g}_{\mu \nu}$, and with spacetime dependent velocities ${\tilde u}_\mu$ and temperatures ${\tilde T}=\frac{1}{\pi {\tilde b}}=e^{-\chi}T$ for arbitrary $\chi(x^\mu)$. 

The fact that different slicings of the same supergravity solution may in fact be interpreted as states
of the same theory in distinct though Weyl equivalent background metrics, reflects the conformal invariance of the dual field theory. As we have seen above, the temperature and velocity fields 
$T$, and $u_\mu$ transform homogeneously under Weyl transformations, with weights $+1$ and $-1$ respectively.

\subsection{Long Wavelength solutions with slowly varying background fields}

In this subsection,we will explain the procedure that we use to 
construct a class of solutions to the Einstein-dilaton action \eqref{eindil} 
that asymptote at large $r$ to the metric and dilaton
\begin{equation}\label{asympn2} \begin{split} 
ds^2 & = -2 u_\mu(x^\mu) dx^\mu dr + r^2 g_{\mu\nu}(x^\mu) dx^\mu dx^\nu \\
\Phi & = \phi(x) \\
\end{split}
\end{equation}
for arbitrary long wavelength functions $\phi(x^\mu)$ and $g_{\mu\nu}(x^\mu)$.
The solutions we construct are dual to fluid dynamics in a field 
theory with an arbitrary long wavelength background metric and coupling 
constant given by $g_{\mu\nu}(x^\mu)$ and $\phi(x^\mu)$. The method we employ 
in this paper closely follows the method in \cite{Bhattacharyya:2008jc}.

\subsubsection{The Zeroth Order ansatz}

We first note that the ansatz 
\begin{equation}\label{genbbn2} \begin{split}
ds^2& = -2 u_\mu(x^\mu) dx^\mu dr + r^2 f(b(x^\mu)r) u_\mu u_\nu dx^\mu dx^\nu 
+ {\cal P}_{\mu\nu} dx^\mu dx^\nu \\
\Phi&=\phi(x^\mu) \\
{\cal P}_{\mu\nu}&=g_{\mu\nu}(x^\mu)+u_\mu(x^\mu) u_\nu(x^\mu) \\
f(r)&=1-\frac{1}{r^4}\\
\end{split}
\end{equation}
for arbitrary functions $u^\mu(x^\mu)$ and $b(x^\mu)$ 
has several attractive features. First, it asymptotes at large $r$ to 
\eqref{asympn1}. Second, all singularities in \eqref{genbbn2} (which 
occur at $r=0$) are shielded from the boundary by a regular event horizon. 
Third, this configuration solves the Einstein-dilaton equations of motion 
for constant $g_{\mu\nu}$, $\phi$, $u_\mu$ and $b$. 

\subsubsection{Setting up a derivative expansion}

As in \cite{Bhattacharyya:2008jc}, we will use \eqref{genbb} as the first term in a systematic derivative expansion of a solution to Einstein's equations. The small parameter that justifies 
this procedure is the inverse length scale of variation of the functions $g_{\mu\nu}$, $\phi$, $u_\mu$ and $b$ (each of which is assumed to vary on the same scale) times the function $b$. Denoting this parameter by $\epsilon$ (see \cite{Bhattacharyya:2008jc} for more details) we plug the expansion 
\begin{equation}\label{pertform} \begin{split}
G_{\mu\nu}&=\sum_{m=0}^\infty \epsilon^m G_{\mu\nu}^{m} \\
\Phi &=\sum_{m=0}^\infty \epsilon^m \Phi^{m} \\
\end{split}
\end{equation}
into the Einstein-dilaton system and solve Einstein's equations perturbatively in $\epsilon$. As we will see, the expansion of these equations to $m^{th}$ order in $\epsilon$ will allow us to determine the functions $G_{\mu\nu}^{m}$ and $\Phi^{m}$. 

\subsubsection{Gauge Choice and Physical Interpretation}

Following \cite{Bhattacharyya:2008jc}, in the bulk of this paper 
we work with the gauge choice 
\begin{equation}\label{gauge}
G_{rr}= 0,\qquad G_{r\mu} \propto u_\mu, \qquad Tr\left( (G^0)^{-1} G^{m} \right) =0 \qquad (m>0)
\end{equation}
The first two of these gauge conditions have been physically interpreted in \cite{Bhattacharyya:2008xc}. As explained in that paper, lines of constant $x^\mu$ are geodesics in any metric subject to this 
gauge condition. The last gauge condition in \eqref{gauge} was chosen arbitrarily and turns out to have no interesting geometrical consequence. However, it is possible to choose a more natural gauge condition with a good geometric interpretation -
\begin{equation}\label{gauget}
G_{rr}=0, \qquad  {G_{r\mu}}=u_\mu
\end{equation}
\eqref{gauget} continues to ensure that lines of constant $x^\mu$ are geodesics  but also 
ensures that the coordinate $r$ is an affine parameter along these geodesics. 

In order to permit easy comparison with the results of \cite{Bhattacharyya:2008jc}, we will work with gauge \eqref{gauge}the bulk of this paper. However, we will also briefly indicate how our final 
results may be transformed into \eqref{gauget}, which will prove useful when comparing with explicit black hole solutions.

\subsubsection{Constraint Equations}

As in \cite{Bhattacharyya:2008jc} it turns out that only 15 of the 16 Einstein
plus dilaton equations are actually independent. These 15 equations may be 
separated into 4 constraint equations (see \cite{Bhattacharyya:2008jc}) and 11 dynamical equations. 

In Appendix \ref{app:forcing}, we show from a direct analysis of the Einstein-dilaton system near the 
boundary that the four constraint equations take the form 
\begin{equation}\label{const}
\nabla_\mu T^{\mu\nu}=  e^{-\phi}\ {\cal L} \nabla^\nu\phi
\end{equation}
where the stress tensor $T_{\mu\nu}$ and Lagrangian ${\cal L}$ 
are defined by in terms of gravity data by the usual formulas of 
AdS/CFT\footnote{Note that the counterterm subtractions 
to the stress tensor and Lagrangian respectively in \eqref{stetc} are simply 
the derivative with respect to $g_{\mu\nu}$ and $\phi$ of 
the counterterm action \cite{Kraus:1999di} 
\begin{equation}\label{acctr}
S_{ctr}=\int \sqrt{h}\left( \frac{R}{2} -\frac{1}{4}(\partial \phi)^2 \right)
\end{equation} 
} 
\begin{equation}\label{stetc}
\begin{split}
16\pi G_5 T^\mu_\nu &= \lim_{r\rightarrow\infty} r^4\left(2(K_{\alpha\beta}h^{\alpha\beta} \delta^\mu_\nu - K^\mu_\nu)\right.\\
&\left.+\bar{\mathcal{G}}^\mu_\nu-6 \delta^\mu_\nu  -\frac{1}{2}\left(\bar{\nabla}^\mu \Phi\bar{\nabla}_\nu\Phi-\frac{\delta^\mu_\nu}{2} (\bar{\nabla} \Phi)^2  \right) \right)\\
16\pi G_5 e^{-\phi}\ {\cal L} &= - \lim_{r\rightarrow\infty} r^4\left(\partial_n\Phi+\frac{1}{2}\bar{\nabla}^2\Phi\right)
\end{split}
\end{equation}
where $n^\mu$ is its outward pointing unit normal of the regulated boundary and $h_{\mu\nu}$ is its induced metric  which leads to the covariant derivative $\bar{\nabla}$ and the corresponding Einstein tensor $\bar{\mathcal{G}}^\mu_\nu$. The extrinsic curvature of the regulated boundary is defined via the normal lie-derivative of the induced metric -  $K_{\mu\nu}\equiv \frac{1}{2} \mathfrak{L}_n h_{\mu\nu} $ and $\partial_n$ is the partial derivative along $n^\mu$. All the indices in the above formulas are raised using the induced metric.

As each term in these equations has an explicit boundary derivative, 
the constraint equations are special from the viewpoint of the boundary derivative 
expansion. This is because each boundary derivative pulls down an additional 
power of $\epsilon$; consequently in the expansion of the constraint equations 
to order $m$ we find contributions from the functions $G_{\mu\nu}^m$ and $\Phi^m$ only 
for $n \leq (m-1)$. Consequently, the constraint equations at $m^{th}$ order do not aid in determining the unknown functions $G_{\mu\nu}^m$ and $\Phi^m$; they instead impose a constraint
on the solution obtained upto $(m-1)^{th}$ order. As this solution has already been determined (by the perturbation theory to $(m-1)^{th}$ order) in terms of $u^\mu(x^\mu)$ and $T(x^\mu)$, these equations effectively reduce to equations of motion for these fluid dynamical fields. In other words, at 
any given order in perturbation theory, the constraint equations are simply 
the equations of fluid dynamics. 

\subsubsection{Dynamical Equations}

The expansion of the remaining 11 independent dynamical equations, to order 
$\epsilon^m$,  yields equations that may be used to determine 
the unknown functions $G^{m}_{\mu\nu}$, $\Phi^{m}$.

The nature of the resultant equations is described in 
\cite{Bhattacharyya:2008jc}. These equations are 
ultralocal in the variables $x^\mu$ (they only contain derivatives of 
$r$). Consequently they may be solved independently 
point by point in $x^\mu$ and are linear ordinary differential equations 
in the variable $r$ at each $x^\mu$. At any given $x^\mu$ these equations 
take the form 
\begin{equation}\label{eqstruct}\begin{split}
M(G^m_{\mu\nu})&=s_{\mu\nu}^m\\
N(\Phi^m)&=s^m\\
\end{split}
\end{equation}
where the homogeneous operators $M$ and $N$ are independent of $m$ 
(they are the same at every order in 
perturbation theory). 

\subsubsection{The Homogeneous Operators}

Let us describe the differential operators $M$ and $N$ in more detail. 
Focus on the equations \eqref{eqstruct} at a 
particular field theory point $y^\mu$. It turns out that the operators 
$M$ and $N$ are simply the equation for linearized 
radial fluctuations of the metric and dilaton in the background of the uniform 
brane \eqref{genbb} with {\it constant} boundary metric $g_{\mu\nu}(y^\mu)$, 
{\it constant} dilaton $\phi(y^\mu) $, {\it constant} velocity $u_\mu(y^\mu)$ 
and {\it constant} inverse temperature $b(y^\mu)$.  In other words the 
homogeneous operators $M$ and $N$ are simply the operators that act on
linearized radial fluctuations of the metric about the background 
\eqref{genbb}.

We can now perform a local co-ordinate transformation to go to Riemann normal co-ordinates where the metric $g_{\mu\nu} = \eta_{\mu \nu}$. We still have some more co-ordinate freedom parameterized by the set of  Lorentz transformations which preserve this form
of the metric.These boost transformations may be used to set $u^\mu =(1,0,0,0)$.
Finally, a scale transformation $ x^\mu \rightarrow \lambda x^\mu$ and 
$r \rightarrow \frac{r}{\lambda}$ may be used to set $b$ to unity. 
Making this choice of coordinates for any given point $y^\mu$ we 
see background metric \eqref{genbb} turns into the metric of a uniform 
black brane at rest with constant temperature $T=\frac{1}{\pi}$ with the 
usual flat boundary metric. However this was precisely the background metric
encountered in the perturbative procedure described in 
\cite{Bhattacharyya:2008jc}. It follows immediately that 
operator $M$ is identical to the operator $H$ in equation 3.4 of 
\cite{Bhattacharyya:2008jc}.

In the same coordinates the operator $N$ is easily determined; it is 
simply given by 
\begin{equation}\label{N}
N=\frac{1}{r^3}\partial_r \left( r^5(1-\frac{1}{r^4})\partial_r \right),
\end{equation}
the radial part of the minimally coupled equation in a black brane at $b=1$.
  
As was explained in \cite{Bhattacharyya:2008jc}, 
the equation $M G_{\mu\nu}=s_{\mu\nu}$ may be solved
by integration for an arbitrary source $s_{\mu\nu}$. As we will explain below, 
the same is true for the operator $N$. Consequently dynamical equations 
may be solved by integration at each order. As in \cite{Bhattacharyya:2008jc}, 
in this paper we will solve these equations subject to two boundary conditions 
\begin{enumerate}
\item $G^m_{\mu\nu}$ and $\Phi^m$ are well behaved (analytic) away from 
the $r=0$ singularity.
\item $\Phi^m$ and $r^2 G^m_{\mu\nu}$ each go to zero as $r \to \infty$. 
\end{enumerate}
The first condition requires no explanation, while the second one ensures 
that our corrections to the metric and dilaton fields do not alter the 
asymptotic form \eqref{asympn2} of our full series solution.

\subsubsection{Source Terms}

We now turn to a brief discussion of the source terms 
$s^m_{\mu\nu}$ and $s^m$. 
As in \cite{Bhattacharyya:2008jc} each of these source terms is determined in terms of lower 
order solutions in perturbation theory, and so may be expressed 
as local functions of $g_{\mu\nu}(x), \phi(x), u^\mu(x)$ and $b(x)$.  
$s^m_{\mu\nu}$ and $s^m$ are each of $m^{th}$ order in derivatives of 
these quantities. The 
dependence of $s^m _{\mu\nu}$ on derivatives of $b$ and $u^\mu$ was already 
described (and determined to second order)  in \cite{Bhattacharyya:2008jc}. 
Below we will also determine the dependence of $s_{\mu\nu}^m$ on derivatives 
of the metric and dilaton (for $m\leq 2$) and compute $s^m$ for $m \leq 2$. 
In this subsubsection we explain the general structure of our results. 

Let us first study the dependence of source terms at $y^\mu$ on derivatives 
of the metric, at the same point,  for $m \leq 2$. For the purpose of this 
calculation we find it useful to work with Riemann normal coordinates 
centered about the point $y^\mu$ in the boundary field theory directions. 
In these coordinates the expansion of the boundary metric $g_{\mu\nu}$ 
about $y^\mu$ starts at second order. As a consequence $s_{\mu\nu}^1$ and $s^1$ each 
are independent of derivatives of $g_{\mu\nu}$. 
While $s_{\mu\nu}^2$ receives contributions proportional to 
the boundary curvature tensor. $s^2$ (like all dilaton source
terms) is necessarily proportional to a derivative of some order of the 
dilaton field \footnote{This follows from the fact that Einstein gravity is a 
consistent truncation of the Einstein-dilaton system.} 
Consequently $s^3$ is the first dilaton source term that receives curvature
dependent contributions; $s^2$ is independent of boundary curvatures. 

We now turn to the dependence of $s_{\mu\nu}^m$ and $s^m$ on derivatives 
of the boundary value of the dilaton for $m \leq 2$. Let us first focus 
on $s^m$. As we have argued each of $s^m$ must be proportional to at least 
one derivative of the boundary dilaton. From symmetry it then follows that 
$s^1 \propto u^\mu \partial \phi$. On the other hand $s^2$ can and does 
receive contributions proportional to both 
$u^\mu u^\nu \partial_\mu \partial_\nu \phi$ as well as terms proportional 
to one derivative of a velocity contracted with a derivative
of $\phi$.

We now discuss the contribution of derivatives of the boundary value of the 
dilaton, $\phi(x)$ to  $s^{m}_{\mu\nu}$.  It follows  
from \eqref{eom}, the dilaton source term for the metric equation is 
proportional to two derivatives of the dilaton. As the dilaton is constant 
(has no $r$ dependence) in the uniform brane solution, and (as we have 
seen in the previous paragraph) is proportional to a derivative of the 
boundary value of $\phi$ at first order, it follows that dilaton 
contributions to $s^{1}_{\mu\nu}$ vanishes, and the contributions of the 
dilaton to $s^2_{\mu\nu}$ are schematically proportional to 
$\partial_\mu \phi \partial_\nu \phi$.

\subsubsection{The Stress Tensor, Lagrangian and Fluid Equations of motion.}

As we have explained above, the dynamical equations upto $m^{th}$ order 
in $\epsilon$ may be used to solve for the metric and dilaton, and 
consequently the stress tensor and the Lagrangian (from \eqref{stetc})
to the same order. By plugging into \eqref{const}, this information also 
determines the fluid dynamical equation (including 
the forcing functions) with $(m+1)$ or fewer derivatives.

This concludes our brief review of the general structure of the long 
wavelength perturbation theory we perform in this paper. We refer the 
reader to \cite{Bhattacharyya:2008jc} for a fuller description of the 
procedure. 

\section{Explicit Results Upto Second Order }\label{sec:exact}

In this section we present explicit formulas for the metric and dilaton, 
as a function of boundary metric, dilaton, velocity and temperature fields, 
to second order in the boundary derivative expansion. We have obtained these 
results by implementing the perturbative procedure described in the previous
section and in \cite{Bhattacharyya:2008jc}. 

\subsection{The metric and dilaton at first order}

As we have described in the previous section, the metric dual to fluid dynamics at first order is already completely determined by the results of 
\cite{Bhattacharyya:2008jc}. $s^1_{\mu\nu}(y^\mu)$ receives no contribution 
from first derivatives of the boundary metric or boundary dilaton  
at $y^\mu$, and the metric $G^1_{\mu\nu}$ is simply by the naive 
boundary covariantization of equation of equation 4.24 of 
\cite{Bhattacharyya:2008jc}.

The dilaton field $\Phi^1$ is nonzero and requires a new - though very simple 
- calculation to determine. 

Using coordinates in which $g_{\mu\nu}(y^\mu)=\eta_{\mu\nu}$, $u_\mu=(-1,1,1,1)$ 
and $b(y^\mu)=1$, the equation for $\Phi^1(y^\mu)$ is 
\begin{equation}\label{dileq}
\partial_r \left(r^5 (1-\frac{1}{r^4} )  \partial_r \Phi^1\right)+
\partial_r (r^3 \partial_v \phi)=0
\end{equation}
This equation may be integrated trivially. The arbitrary solution to this 
differential equation is given by
\begin{equation}\label{arbsol}
\Phi^{1}=c_1(x^\mu) + c_2 \int_{r}^\infty \frac{1}{r^5 f(r)} + 
\partial_v \phi \int_r^\infty \frac{r^3-1}{r^5 f(r)} 
\end{equation}
Our boundary condition at infinity forces $c_1=0$ while the requirement 
of regularity at the horizon $r=1$ sets $c_2=0$\footnote{Note that, for any non-zero value
of $c_2$, the dilaton has a logarithmic divergence at $r=1$. This (co-ordinate invariant) singularity is not shielded by any horizon and hence we will consider it to be physically unacceptable. We will show in the later sections that when $c_2=0$, our solutions have regular event horizon that passes through the neighbourhood of $r=1$.  } . Consequently we conclude 
\begin{equation}\label{acsol}
\Phi^{1}= 
\partial_v \phi \int_r^\infty \frac{r^3-1}{r^5 f(r)} 
=u.\partial \phi  \int_r^\infty \frac{r^3-1}{r^5 f(r)}
\end{equation}

\subsection{Solution at second order}

The computations required to determine the metric and dilaton field at 
second order, while involved in practice, are a straightforward 
generalization of the calculations presented in \cite{Bhattacharyya:2008jc}. 
We have performed these calculations with the aid of the symbolic manipulation 
program Mathematica. In this section we simply record our final results.

The metric upto second order is given by 
\begin{equation}\label{dwitio}
\begin{split}
ds^2 =& -2 u_\mu(x^\mu) dx^\mu dr + r^2 f(b(x^\mu)r) u_\mu u_\nu dx^\mu dx^\nu 
+ {\cal P}_{\mu\nu} dx^\mu dx^\nu \\
+& \left(2\ b\ r^2 F(b r) \sigma_{\mu\nu} + \frac{2}{3} r\ \theta\ u_\mu u_\nu -r (a_\mu u_\nu + a_\nu u_\mu)\right)dx^\mu dx^\nu\\
+& 3\  b^2 H\ u_{\mu} dx^\mu dr\\
 +& \left(r^2 b^2 H\ {\cal P}_{\mu\nu} + \frac{1}{r^2 b^2}K\ u_\mu u_\nu + \frac{1}{r^2 b^2}\left(J_\mu u_\nu + J_\nu u_\mu\right) + r^2 b^2 \alpha_{\mu\nu}\right)dx^\mu dx^\nu\\
\Phi& = \phi(x^\mu) + u.\partial \phi \int_{rb}^\infty dx  \frac{x^3-1}{x^5 f(x)} 
+ {\mathcal S}_\phi^{(1)} h_\phi^{(1)}(br) + {\mathcal S}_\phi^{(2)} h_\phi^{(2)}(br) \\
\end{split}
\end{equation}
In this equation the first line is simply the ansatz \eqref{genbbn2}. 
The second line records corrections to this metric at first order in 
$\epsilon$, while the third and the fourth lines record the second order 
corrections to this ansatz.  

\eqref{dwitio} - the metric dual to fluid flows upto second order in the 
derivative expansion - is one of the main results of this paper. In the rest of
this section we will systematically define all the previously undefined 
functions functions that appear in \eqref{dwitio}. We will start by defining 
all scalar functions of the radial coordinate $r$ that appear in 
\eqref{genbbn2}, and then turn to the definition of the index valued 
forms that these functions multiply. 

The only undefined function of $r$ in the second line of 
\eqref{dwitio} is $F(r)$ which is given by\footnote{As explained in \cite{Bhattacharyya:2008jc}, we obtain $F(r)$ by integrating a second-order
differential equation. One of the integration constants of this equation are fixed the requirement of regularity of $F(r)$ at $r=1$. For a general value of this constant,
$F(r)$ has a logarithmic singularity at $r=1$. This singularity is physical rather than 
a coordinate artifact, and is hence unacceptable. We have checked the last statement by computing the curvature invariant $R_{MN}R^{MN}$; this quantity in general has a pole type singularity, proportional to $\sigma_{\mu\nu}\sigma^{\mu\nu}$ at $r=1$. }
\begin{equation}\label{prothom}
F(r) = {\frac{1}{4}}\, \left[\ln\left(\frac{(1+r)^2(1+r^2)}{r^4}\right) - 2\,\arctan(r) +\pi\right] 
\end{equation}
The undefined functions on the third fourth and fifth 
line of the same equation are defined as  
\begin{equation}\label{duisc}
\begin{split}
H &= h^{(1)}(b r)\ \ST{4} + h^{(2)}(b r)\ \ST{5} + h^{(3)}(br) \ \ST{1}^\phi\\
K &= k^{(1)}(b r)\ \ST{4} + k^{(2)}(b r)\ \ST{5} + k^{(3)}(b r)\ {\cal S}
+ k^{(4)}(br) \ST{1}^\phi + k^{(5)}(br) \ST{2}^\phi\\
J_\mu &= j^{(1)}(b r)\  \mathbf{B}^{\infty}_{\mu} + j^{(2)}(b r)\  \mathbf{B}^{\text{fin}}_{\mu} +j^{(3)}(br) {\bf B}^\phi_\mu \\
\alpha_{\mu\nu} &= a_1(br) {\cal T}_{\mu\nu} + a_5(br)\  (T_5)_{\mu\nu}\\
 &+ a_6(br)\  (T_6)_{\mu\nu} + a_7(br)\  (T_7)_{\mu\nu} +a_8(br)\ C_{\mu\nu} 
+a_9(br) (T_\phi)_{\mu\nu}
\end{split}
\end{equation}
where
\begin{equation}\label{duifun}
\begin{split}
h^{(1)}(r) &= -\frac{1}{12 r^2}\\
h^{(2)}(r) &=  -\frac{1}{6 r^2} + \int_r^\infty \frac{dx}{x^5}\int_x^\infty dy\  y^4\left(\frac{1}{2}W_h(y) - \frac{2}{3 y^3}\right)\\
h^{(3)}(r)&=\frac{r^4+3}{96 r^4} \pi  -\frac{r^4+3}{48 r^4} \tan^{-1}(r)
-\frac{1}{12} \ln r \\
&+\frac{r^4+3}{48 r^4} \ln(1+r) +\frac{r^4-1}{32 r^4}\ln(1+r^2)
+\frac{-2 r^3+3 r^2+1}{48 r^4}\\
\end{split}
\end{equation}
\begin{equation}
\begin{split}
k^{(1)}(r) &= -\frac{r^2}{12} -\int_r^\infty \left(12 x^3 h^{(1)}(x) + (3 x^4 -1)\frac{d h^{(1)}(x)}{dx} + \frac{1 + 2 x^4}{6 x^3} + \frac{x}{6}\right) \\
k^{(2)}(r) &= \frac{7 r^2}{6} -  \int_r^\infty \left(12 x^3 h^{(2)}(x) + (3 x^4 -1)\frac{d h^{(2)}(x)}{dx} + \frac{1}{2}W_k(x) - \frac{7 x}{3}\right)\\
k^{(3)}(r) &= r^2/2\\
k^{(4)}(r) &=\frac{r^8-1}{32 r^4} \pi -\frac{r^8-1}{16 r^4} \tan^{-1}(r)
+ \frac{1}{4} \left(1-r^4\right) \ln r\\
& + \frac{r^8-1}{16 r^4} \ln(1+r)
+ \frac{3 r^8-4 r^4+1}{32 r^4} \ln(1+r^2) \\ 
&+ \frac{-6 r^7+5 r^6+3 r^4+2 r^3-3 r^2-1}{48 r^4}\\
k^{(5)}(r) &= r^2/12 \\
\end{split}
\end{equation}
\begin{equation}
\begin{split}
j^{(1)}(r) &= \frac{r^2}{36} -\int_r^\infty dx\ x^3 \int_x^\infty dy \left(\frac{p(y)}{18 y^3 (y+1)(y^2 + 1)} -\frac{1}{9 y^3}\right) \\ 
j^{(2)}(r) &= -\int_r^\infty dx\ x^3 \int_x^\infty dy\left(\frac{1}{18 y^3 (y+1)(y^2 + 1)}\right)\\
j^{(3)}(r) &=-\frac{1-r^4}{16}\left( \pi - 2\tan^{-1}(r) +2 \ln(1+r) -\ln(1+r^2) \right) 
-\frac{r^2(1+2r)}{8}\\
\end{split}
\end{equation}
\begin{equation}
\begin{split}
a_1(r) &= -\int_r^\infty\frac{dx}{x(x^4 - 1)}\int_1^x dy\ 2 y\left(\left[\frac{3 p(y) + 11}{p(y) + 5}\right] - 3 y F(y)\right)\\
a_5(r) &= -\int_r^\infty\frac{dx}{x(x^4 - 1)}\int_1^x dy\ y\left(1 + \frac{1}{y^4}\right)\\
a_6(r) &= -\int_r^\infty\frac{dx}{x(x^4 - 1)}\int_1^x dy\ 2y \left(\frac{4}{y^2}\left[\frac{y^2p(y) + 3y^2 -y -1}{p(y) + 5}\right] - 6 y F(y)\right)\\
a_7(r) & = \frac{1}{4}\int_r^\infty\frac{dx}{x(x^4 - 1)}\int_1^x dy\ 2y \left(2\left[\frac{p(y) + 1}{p(y) + 5}\right] - 6 y F(y)\right)\\
a_8(r) &= -\int_r^\infty\frac{dx}{x(x^4 - 1)}\int_1^x dy\ 2y\\
a_9(r) &= \frac{\ln(1+\frac{1}{r^2} )}{4}\\ 
\end{split}
\end{equation}
\begin{equation}
\begin{split}
W_h(r) &= \frac{4}{3}\frac{(r^2 + r + 1)^2 - 2 (3r^2 + 2r + 1)F(r)}{r(r+1)^2(r^2+1)^2}\\
W_k(r) &= \frac{2}{3}\frac{4(r^2 + r + 1)(3r^4 -1)F(r)- (2r^5 + 2r^4 + 2r^3 - r - 1)}{r(r+1)(r^2 + 1)}\\
p(r) &= 2r^3 + 2r^2 + 2r - 3\\
h_\phi^{(1)}(r) &= -\int_r^\infty  \frac{dx}{x(x^4 - 1)}\int_1^x dy\left[-\frac{y^2}{4}\ln\left[(1 + y)(1+  y^2 )\right] \right.\\
&\left.+ \frac{y^2}{2}\tan^{-1}(y) + \frac{y(1 + y )( 1 + y^2)-2 y}{3(1 + y)(1 + y^2)}\right]\\
h_\phi^{(2)}(r) &= \int_r^\infty \frac{dx}{x(x^4 - 1)}\int_1^x dy\left(\frac{2y}{3}\right)
\end{split}
\end{equation}

We now turn to defining all the terms that carry boundary index structure in 
\eqref{genbbn2}. These terms are all expressed in terms of fixed numbers 
of boundary derivatives of the velocity, metric and boundary dilaton fields. 

{\it Terms with a single boundary derivative}
\begin{equation}\label{onederi}
\begin{split}
\theta &= \nabla_\alpha u^\alpha\\
a_\mu &= (u.\nabla)u_\mu\\
l^\mu &= \epsilon^{\alpha\beta\gamma\mu}u_\alpha\nabla_\beta u_\gamma\\
\sigma_{\mu\nu} &= \frac{1}{2}{\cal P}^{\mu\alpha}{\cal P}^{\nu\beta}\left(\nabla_\alpha u_\beta + \nabla_\beta u_\alpha\right) - \frac{1}{3}{\cal P}_{\mu\nu}\theta\\
\end{split}
\end{equation}
The quantity $l_\mu$ defined here does not appear in the first order 
correction to the ansatz metric, but does appear, multiplied by 
other first order terms appears in the second order metric.
\\
We now describe all terms with two boundary derivatives. Following 
\cite{Bhattacharyya:2008jc} we sub-classify these terms as scalar like, vector
like or tensor like, depending on their transformation properties under the 
$SO(3)$ rotation group that is left unbroken by the velocity $u_\mu$ 
(see \cite{Bhattacharyya:2008jc} for more details) 
\\
{\it Scalar terms with two derivatives}
\begin{equation}\label{scaltwoderi}
\begin{split}
{\cal S} &= \left(-\frac{4}{3}(\Ss{3} - R_1) + 2\ \ST{1} -\frac{2}{9}\ST{3}\right)\\
\ST2 &= l_\mu a^\mu\\
\ST4 &= l_\mu l^\mu\\
\ST5 &= \sigma_{\mu\nu}\ \sigma^{\mu\nu}\\
\ST1^\phi&=u^\mu u^\nu \nabla_\mu \phi \nabla_\nu \phi\\
\ST2^\phi&= {\cal P}^{\mu\nu} \nabla_\mu \phi \nabla_\nu \phi\\
{\mathcal S}^{(1)}_\phi&= 3 u^\mu u^\nu \nabla_\mu  \nabla_\nu \phi + 3 a^\mu\nabla_\mu \phi + \theta u^\mu\nabla_\mu\phi\\
{\mathcal S}^{(2)}_\phi&= \frac{3}{2}\nabla^2 \phi - 3 a^\mu\nabla_\mu \phi + \theta u^\mu\nabla_\mu\phi\\
\end{split}
\end{equation}
where
\begin{equation}\label{exscal}
\begin{split}
\ST1 &= a_\mu a^\mu\\
\ST3 &= \theta^2\\
\Ss{3} &= \frac{1}{b}{\cal P}^{\alpha\beta}\nabla_{\alpha}\nabla_{\beta}\ b\\
R_1 &= -\frac{1}{4}{\cal P}^{\alpha\beta}{\cal P}^{\gamma\nu}R_{\alpha\gamma\beta\nu}\\
\end{split}
\end{equation}
\\
{\it Vector terms with two derivatives}
\begin{equation}\label{vectwoderi}
\begin{split}
{\bf B}^\infty  &= 4\, \left(10\, \V{4} +  \V{5} +3\, \VT{1} -3 \,\VT{2} -6\,\VT{3} + 9\ R_2 \right)\\
{\bf B}^{{\rm fin}} & = 9\, \left(20\,  \V{4}- 5 \,\VT{2} - 6\,\VT{3} + 20\ R_2\right) \\
{\bf B}^\phi_\mu &=u.\nabla \phi ~{\cal P}_\mu^\nu \nabla_\nu \phi\\ 
\end{split}
\end{equation}
where
\begin{equation}\label{exvec}
\begin{split}
(\V{4})_\nu &= \frac{9}{5}\left[\frac{1}{2}{\cal P}^\alpha_\nu{\cal P}^{\beta\gamma}\left(\nabla_\beta u_\gamma + \nabla_\gamma u_\beta\right) - \frac{1}{3}{\cal P}^{\alpha\beta}{\cal P}^\gamma_\nu\nabla_\gamma\nabla_\alpha\ u_\beta \right] -{\cal P}^{\alpha\beta}{\cal P}_\nu^\gamma\nabla_\alpha\nabla_\beta\ u_\gamma\\
(\V{5})_\nu &= {\cal P}^{\alpha\beta}{\cal P}_\nu^\gamma\nabla_\alpha\nabla_\beta\ u_\gamma\\
\VT{1}_\nu &= a_\nu\ \theta\\
\VT{2}_\nu &= \epsilon_{\alpha\beta\gamma\nu}u^\alpha\ a^\beta\ l^\gamma\\
\VT{3}_\nu &= a^\alpha\ \sigma_{\alpha\nu}\\
(R_2)_\nu &=  -\frac{1}{2}{\cal P}^{\alpha\beta}{\cal P}^\gamma_\nu\ u^\mu\ R_{\mu\alpha\gamma\beta}\\
\end{split}
\end{equation}
\\
{\it Tensor terms with two derivatives}
\begin{equation}\label{tentewoderi}
\begin{split}
\mathfrak{T}_{\mu\nu}  &= (\mathfrak{T}_1)_{\mu\nu} + \frac{1}{3} (\mathfrak{T}_4)_{\mu\nu} + 
(\mathfrak{T}_3)_{\mu\nu}\\
(\mathfrak{T}_{5})_{\mu\nu} &= l_\mu l_\nu - \frac{1}{3}{\cal P}_{\mu\nu}\ \ST{4}\\
(\mathfrak{T}_{6})_{\mu\nu} &= \sigma_{\mu\alpha}\ \sigma_\nu^\alpha - \frac{1}{3}{\cal P}_{\mu\nu}\ \ST{5}\\
(\mathfrak{T}_{7})_{\mu\nu} &= \left(\epsilon^{\alpha\beta\gamma\mu}\ \sigma^\nu_\gamma + \epsilon^{\alpha\beta\gamma\nu}\ \sigma^\mu_\gamma\right)u_\alpha\ l_\beta\\
C_{\mu\nu} &= {\cal P}_\mu^\alpha{\cal P}_\nu^\beta\left(R_{\alpha\beta} + 2 u^\gamma\ u^\lambda R_{\gamma\alpha\lambda\beta}\right) - \frac{1}{3}{\cal P}_{\mu\nu}{\cal P}^{\alpha\beta}\left(R_{\alpha\beta} + 2 u^\gamma\ u^\lambda R_{\gamma\alpha\lambda\beta}\right) \\
(T_\phi)_{\mu\nu}&={\cal P}_\mu^\alpha {\cal P}_\nu^\beta ~\nabla_\alpha \phi~\nabla_\beta \phi 
-\frac{1}{3} {\cal P}_{\mu\nu} \mathcal{P}^{\alpha\beta}(\partial_\alpha \phi)(\partial_\beta \phi) 
\end{split}
\end{equation}
where
\begin{equation}\label{exten}
\begin{split}
(\mathfrak{T_1})_{\mu\nu} &= a_\mu a_\nu - \frac{1}{3}{\cal P}_{\mu\nu}\ \ST{1}\\
(\mathfrak{T_3})_{\mu\nu} &=\frac{1}{2}{\cal P}_\mu^\alpha{\cal P}_\nu^\beta\ (u.\nabla)\left(\nabla_\alpha u_\beta + \nabla_\beta u_\alpha\right) -\frac{1}{3}{\cal P_{\mu\nu}}{\cal P}^{\alpha\beta}(u.\nabla)(\nabla_\alpha u_\beta)\\
(\mathfrak{T_4})_{\mu\nu} &= \sigma_{\mu\nu}\theta\\
\end{split}
\end{equation}

Here $\nabla$ denotes the covariant derivative in the curved boundary metric, $R_{\alpha\beta\gamma\nu}$ is the Riemann tensor and $R_{\mu\nu}$ is the Ricci tensor.

The metric and dilaton configuration presented in this subsection solves 
the coupled Einstein-dilaton equations of motion provided that the temperature
and velocity fields obey the equation of motion presented in the next 
subsection.

\subsection{Shift of Gauge}

While we have (for historical reasons) presented our final metric 
gauge choice \eqref{gauge}, it would have been more natural, in some respects, to work with the gauge choice \eqref{gauget}. The variable change that 
converts from \eqref{gauge} to \eqref{gauget} is simply given, 
to second order, by $d {\tilde r} = d {r}(1-\frac{3}{2} b^2 H(b r) )$. 
This variable change
may be integrated. The metric expressed in terms of the variable ${\tilde r}$ 
will be in the gauge \eqref{gauget} at second order.

\subsection{The Stress Tensor, Lagrangian and Forcing function at second order}

The stress tensor and expectation value of the `Lagrangian' may be 
determined for the field theory configurations dual to the solutions 
of the previous subsection using the formulas \eqref{stetc}. We find 
\begin{equation}\label{stetcn} \begin{split}
16 \pi G_5 T^{\mu\nu}&=  (\pi \,T)^4
\left( g^{\mu \nu} +4\, u^\mu u^\nu \right) -2\, (\pi\, T )^3 \,\sigma^{\mu \nu} \\
& + (\pi T)^2 \,\left( \left(\frac{\ln 2}{2}\right) \, (\mathfrak{T}_7)^{\mu\nu} +2\, 
(\mathfrak{T}_{6})^{\mu\nu} +
\left( 2- \ln2 \right)  \mathfrak{T}^{\mu\nu}  + C^{\mu \nu} -
\frac{1}{2} T_{\phi}^{\mu\nu}\right) \\
-16 \pi G_5 e^{-\phi}{\cal L}&=(\pi T)^3 ~u.\partial \phi +  (\pi T)^2\left(\frac{1}{3} {\cal S}_\phi^{(2)} + \frac{\ln 2}{6}{\cal S}_\phi^{(1)}\right)
\end{split}
\end{equation}

Further, the spacetime configuration presented in the previous section is 
a solution to the equations of motion if and only if the velocity and 
temperature fields obey the constraint 
\begin{equation}\label{enmomforce}\begin{split}
\nabla_\mu T^{\mu \nu}& = f^\nu \\
f^\nu& =e^{-\phi}{\cal L}\nabla^\nu \phi 
\end{split}
\end{equation}

\section{Constraints from Weyl Covariance}\label{sec:weyl}

In the previous section we have determined the bulk metric dual to 
a particular fluid flow $u_\mu(x)$ on a boundary manifold with a particular 
background metric $g_{\mu\nu}(x^\mu)$ and boundary dilaton field $\phi(x^\mu)$.
However, as we have explained in section \ref{sec:deriv:exp}, this must be the same as the 
bulk geometry dual to the velocity field $\tilde{u}_\mu(x^\mu)=
e^\chi u_\mu(x^\mu)$ on the boundary space with metric 
$\tilde{g}_{\mu\nu}(x^\mu)=e^{2 \chi} g_{\mu\nu}(x^\mu)$ and the boundary 
dilaton unchanged. This equality of spaces, which is not algebraically 
automatic in the formulas of our perturbation theory, constitutes a tight
algebraic check on our procedure. 

In this section we will demonstrate that the metric described in the previous
section passes this test. In more detail we will demonstrate that the metrics 
dual to fluid dynamics must are invariant under the simultaneous 
replacements\footnote{It may be useful to reiterate the logic that underlies 
this test.  Let us imagine 
we have solved the problem described in the previous subsection for a given 
background metric $g_{\mu \nu}$ and velocity and temperature fields 
$u_\mu(x)$ and $b(x)$. Upon performing the coordinate transformation 
$r={\tilde r} e^{-\chi(x^\mu)}$ (as described above) for a slowly varying 
function $\chi(x)$ we have a gravitational background of the form \eqref{genbbn2}
with the new metric velocity and temperature functions ${\tilde u}_\mu=
e^{\chi} u_\mu$ and ${\tilde g}_{\mu\nu}=e^{2 \chi}{\tilde g}_{\mu\nu}$ 
and ${\tilde b}=e^{\chi} b$. However we could also have directly solved
for the metric dual to fluid dynamics with this data.}
 
\begin{equation} \label{wt}
{\tilde u}_\mu \rightarrow e^{\chi} u_\mu, ~~~{\tilde g}_{\mu\nu} 
\rightarrow e^{2 \chi}{\tilde g}_{\mu\nu},  ~~~b \rightarrow e^{\chi} 
{\tilde b }, ~~~ r \rightarrow e^{-\chi} {\tilde r}, ~~~ \phi \rightarrow 
\phi 
\end{equation}

This property of our solutions in particular implies the Weyl covariance
of the fluid dynamical stress tensor (and Lagrangian) that follows from 
our solutions. 
 
\subsection{Weyl Covariant Formalism and Independent Weyl Covariant Tensors}\label{subsec:WeylTens}

Note that the transformation \eqref{wt} is simply a Weyl transformation in the 
boundary field theory directions. Consequently, in order to investigate the 
invariance properties of our solutions under the transformation 
\eqref{wt} we find it convenient to employ the manifestly Weyl-covariant 
formalism for hydrodynamics that was developed in \cite{Loganayagam:2008is}. 
The main technical innovation of \cite{Loganayagam:2008is} was the introduction
of a Weyl-covariant derivative, whose action on an arbitrary tensor field 
is defined by 
\begin{equation}\label{D:eq}
\begin{split}
\mathcal{D}_\lambda\ Q^{\mu\ldots}_{\nu\ldots} &\equiv \nabla_\lambda\ Q^{\mu\ldots}_{\nu\ldots} + w\  \mathcal{A}_{\lambda} Q^{\mu\ldots}_{\nu\ldots} \\ 
&+\left[{g}_{\lambda\alpha}\mathcal{A}^{\mu} - \delta^{\mu}_{\lambda}\mathcal{A}_\alpha  - \delta^{\mu}_{\alpha}\mathcal{A}_{\lambda}\right] Q^{\alpha\ldots}_{\nu\ldots} + \ldots\\
&-\left[{g}_{\lambda\nu}\mathcal{A}^{\alpha} - \delta^{\alpha}_{\lambda}\mathcal{A}_\nu  - \delta^{\alpha}_{\nu}\mathcal{A}_{\lambda}\right]  Q^{\mu\ldots}_{\alpha\ldots} - \ldots
\end{split}
\end{equation}
where the Weyl-connection $\mathcal{A}_\mu$ is related to the fluid velocity via the relation
\begin{equation}\label{defA:eq}
\begin{split}
\mathcal{A}_\mu = a_\mu - \frac{\vartheta}{d-1} {u}_\mu
\end{split}
\end{equation}

As a technical prerequisite to our main goal, in the rest of this 
section we employ this Weyl covariant formalism to 
list the most general Weyl invariant scalars, transverse vectors and 
symmetric traceless transverse tensors in hydrodynamics that involve no more than second order derivatives.\footnote{We will restrict attention to fluid dynamics in $3+1$ dimensions.} We perform this listing, taking particular care 
to account for the equations of motion. In other words tensor fields that are 
equivalent on-shell are counted only once in our listing. 

Let us first start by eliminating an easily dealt with redundancy. 
As an arbitrary function of the dilaton is Weyl-invariant, it follows that we 
can generate new Weyl invariants by multiplying old ones by functions of 
$\phi$ to get new Weyl-invariant observables. We shall omit the observables 
formed this way from the lists below, but take care to account for them later. 

To begin, let us start with the basic quantities of hydrodynamics - 
the fluid temperature $T$ and the fluid velocity 
$u^\mu$\footnote{We will assume that there are no other conserved charges except the energy momentum tensor.}  . The former is a Weyl-covariant scalar with conformal weight unity and the latter is a Weyl-covariant vector with conformal weight unity. It follows that, at the zero derivative level, there are no 
non-trivial Weyl-invariant scalars, no transverse vector or  symmetric 
traceless transverse tensors.

\footnote{We shall follow the notations of \cite{Loganayagam:2008is} in the rest of this section(except for the curvature tensors which differ by a sign from the curvature tensors in \cite{Loganayagam:2008is}. In particular, we recall the following definitions 
\begin{equation}\label{weylcov:eq}
\begin{split}
\mathcal{A}_\mu = a_\mu - \frac{\vartheta}{3}u_\mu \ ;\qquad & \mathcal{F}_{\mu\nu} = \nabla_\mu \mathcal{A}_\nu - \nabla_\nu \mathcal{A}_\mu \\
\mathcal{R} = R +6 \nabla_\lambda \mathcal{A}^\lambda - 6 \mathcal{A}_\lambda \mathcal{A}^\lambda \ ; \qquad& \mathcal{D}_\mu u_\nu = \sigma_{\mu\nu} + \omega_{\mu\nu} \\
\mathcal{D}_\lambda \sigma^{\mu\lambda} = \nabla_\lambda \sigma^{\mu\lambda}- 3 \mathcal{A}_\lambda \sigma^{\mu\lambda} \ ;\qquad& \mathcal{D}_\lambda \omega^{\mu\lambda} = \nabla_\lambda \omega^{\mu\lambda}- \mathcal{A}_\lambda \omega^{\mu\lambda}  \\
\end{split}
\end{equation} 
Note that in a flat spacetime, $R$ is zero but $\mathcal{R}$ is not.}

At one derivative level $T^{-1}u^\mu \mathcal{D}_\mu \phi$ is the only Weyl invariant scalar. The only Weyl invariant transverse vector is $P^\nu_\mu \mathcal{D}_\nu \phi$. Finally, the  only Weyl-invariant transverse pseudo-vector $l_\mu$ and only one Weyl-invariant symmetric traceless transverse tensor $T\sigma_{\mu\nu}$.

At the two derivative level, there are seven independent Weyl-invariant scalars
\begin{equation} \label{scal}
\begin{split}
T^{-2}\sigma_{\mu\nu}\sigma^{\mu\nu},\qquad& T^{-2}\omega_{\mu\nu}\omega^{\mu\nu},\qquad  T^{-2}\mathcal{R},\\
T^{-2}P^{\mu\nu}\mathcal{D}_\mu \mathcal{D}_\nu \phi ,\qquad T^{-2}u^{\mu}u^{\nu}\mathcal{D}_\mu \mathcal{D}_\nu \phi,&\qquad T^{-2}P^{\mu\nu}\mathcal{D}_\mu \phi\mathcal{D}_\nu \phi \qquad  \text{and}\qquad T^{-2}u^{\mu}u^{\nu}\mathcal{D}_\mu \phi\mathcal{D}_\nu \phi
\end{split}
\end{equation}
one Weyl-invariant pseudo-scalar $T^{-2}l^{\mu}\mathcal{D}_\mu \phi$ and six independent Weyl-invariant transverse vectors  
\begin{equation*}
\begin{split}
T^{-1}P_\mu^{\nu}\mathcal{D}_\lambda\sigma_{\nu}{}^\lambda, \qquad T^{-1}P_\mu^{\nu}\mathcal{D}_\lambda\omega_{\nu}{}^\lambda, &\qquad T^{-1}P_\mu^{\nu}u^\lambda\mathcal{D}_\nu \mathcal{D}_\lambda \phi,\qquad T^{-1}P_\mu^{\nu}u^\lambda\mathcal{D}_\nu \phi \mathcal{D}_\lambda \phi,\\
\qquad T^{-1}\sigma_\mu{}^{\lambda}  \mathcal{D}_\lambda \phi,\qquad
 &\text{and} \qquad  T^{-1}\omega_\mu{}^{\lambda}  \mathcal{D}_\lambda \phi 
\end{split}
\end{equation*}
and two Weyl-invariant transverse pseudo-vectors $T^{-1}u^\lambda\mathcal{D}_\lambda l_\mu$ and $T^{-1}l_\mu u^\lambda\mathcal{D}_\lambda\phi$ .

There are eight Weyl-invariant symmetric traceless transverse tensors - 
\begin{equation} \label{sym}
\begin{split}
u^\lambda\mathcal{D}_\lambda\sigma_{\mu\nu},\qquad \sigma_{\mu\nu} u^\lambda\mathcal{D}_\lambda\phi ,&\qquad C_{\mu\alpha\nu\beta}u^\alpha u^\beta ,\qquad \omega_{\mu}{}^{\lambda}\sigma_{\lambda\nu}+\omega_{\nu}{}^{\lambda}\sigma_{\lambda\mu},\\
\frac{1}{2}\left[P^\alpha_\mu P^\beta_\nu + P^\alpha_\nu P^\beta_\mu - \frac{2}{3} P^{\alpha\beta}P_{\mu\nu} \right] \mathcal{D}_\alpha \mathcal{D}_\beta \phi ,&\qquad \frac{1}{2}\left[P^\alpha_\mu P^\beta_\nu + P^\alpha_\nu P^\beta_\mu - \frac{2}{3} P^{\alpha\beta}P_{\mu\nu} \right] \mathcal{D}_\alpha \phi\ \mathcal{D}_\beta \phi, \\
\sigma_{\mu}{}^{\lambda}\sigma_{\lambda\nu}-\frac{P_{\mu\nu}}{3}\  \sigma_{\alpha\beta}\sigma^{\alpha\beta}\qquad  &\text{and}\qquad
\omega_{\mu}{}^{\lambda}\omega_{\lambda\nu}+\frac{P_{\mu\nu}}{3}\  \omega_{\alpha\beta}\omega^{\alpha\beta} .\\
\end{split}
\end{equation}
and five Weyl-invariant symmetric traceless transverse pseudo-tensors\footnote{Note that the
last tensor is not really a pseudotensor - but we count it along with the other tensors
which need $\epsilon^{\alpha\beta\lambda\mu}$ for their definition.}
\begin{equation*}
\begin{split}
\mathcal{D}_\mu l_\nu + \mathcal{D}_\nu l_\mu,&\qquad l_\mu P^\lambda_\nu\mathcal{D}_\lambda\phi + l_\nu P^\lambda_\mu\mathcal{D}_\lambda\phi  - \frac{2P_{\mu\nu}}{3} l^\lambda \mathcal{D}_\lambda\phi \\
\epsilon^{\alpha\beta\lambda}{}_\mu C_{\alpha\beta\nu\sigma}u^\sigma u_\lambda +
\epsilon^{\alpha\beta\lambda}{}_\nu C_{\alpha\beta\mu\sigma}u^\sigma u_\lambda,&\qquad \epsilon^{\alpha\beta\lambda}{}_{\mu}u_\alpha\mathcal{D}_\beta\phi\ \sigma_{\lambda\nu} +\epsilon^{\alpha\beta\lambda}{}_{\nu}u_\alpha\mathcal{D}_\beta\phi \sigma_{\lambda\mu}\\
\text{and} \qquad &\frac{1}{4}\ \epsilon^{\alpha\beta\lambda}{}_{\mu}\ \epsilon^{\gamma\theta\sigma}{}_{\nu} C_{\alpha\beta\gamma\theta}\ u_\lambda u_{\sigma} .
\end{split}
\end{equation*}

\subsection{Manifestly Weyl covariant form of the fluid dynamical metric}

In either of the gauges \eqref{gauge} or \eqref{gauget} employed in this paper, a bulk derivative metric that is invariant under the transformations 
\eqref{wt} must be constructed out of
\begin{enumerate}
\item The boundary Weyl invariant tensors listed in the previous 
subsection.
\item The Weyl-invariant vector-valued $1-$form 
$u_\mu(dr+r \mathcal{A}_\lambda dx^\lambda)$
\item Arbitrary functions of  the Weyl-invariant scalar $\xi=r/(\pi T)$
\end{enumerate}.
Consequently the metric must take the form 
\begin{equation}\label{deriv:met:eq}
\begin{split}
ds^2 &= -2 W_1 \ u_\mu dx^\mu(dr+r \mathcal{A}_\lambda dx^\lambda) \\
&+\left[ r^2\left( W_2 \ g_{\mu\nu} + W_3 u_\mu u_\nu  \right) + r\left( W_{4\mu} u_\nu + u_\mu W_{4\nu}\right) + W_{5\mu\nu}  \right] dx^\mu dx^\nu . \\
\end{split}
\end{equation}
where $W_1,W_2$ and $W_3$ are Weyl-invariant scalars, $W_{4\mu}$ is a Weyl-invariant transverse vector and $W_{5\mu\nu}$ is a Weyl-invariant transverse traceless tensor. It follows that the $W_i$'s can be expressed as functions of $\xi$ and the Weyl covariant observables in fluid dynamics with appropriate weights.

In the rest of this section we will check that our metric \eqref{dwitio}
is indeed of this form. We will also compute all the functions that enter
into \eqref{deriv:met:eq}

\subsection{Weyl Covariant form of the second order metric}

In order to cast the metric in this paper in the form given by \eqref{deriv:met:eq} , we have to first rewrite the quantities appearing in the metric in a Weyl-covariant form. We have the following relations in the flat spacetime which identify the Weyl-covariant forms appearing in the second-order metric -
\begin{equation}\label{covforms:eq}
\begin{split}
\mathfrak{S}_1{}^\phi = u^\mu u^\nu \mathcal{D}_\mu\phi\ \mathcal{D}_\nu \phi ;\qquad& \mathfrak{S}_2{}^\phi = P^{\mu\nu} \mathcal{D}_\mu\phi\ \mathcal{D}_\nu \phi \\
\mathfrak{S}_4 = 2 \omega_{\alpha\beta} \omega^{\alpha\beta} ;\qquad \mathfrak{S}_5 = \sigma_{\alpha\beta} \sigma^{\alpha\beta};\qquad& \mathcal{S}^{(1)}_\phi = 3 u^\mu u^\nu \mathcal{D}_\mu \mathcal{D}_\nu \phi ;\qquad  \mathcal{S}^{(2)}_\phi = \frac{3}{2} \mathcal{D}_\mu \mathcal{D}^\mu \phi\\
-\frac{4}{3} (s_3-R_1) + 2 \mathfrak{S}_1 -\frac{2}{9}\mathfrak{S}_3 &= \frac{2}{3} \sigma_{\alpha\beta} \sigma^{\alpha\beta}-\frac{2}{3} \omega_{\alpha\beta} \omega^{\alpha\beta}-\frac{1}{3}\mathcal{R} \\
\frac{5}{9}v_{4\mu} + \frac{5}{9} v_{5\mu} +\frac{5}{3} \mathfrak{V}_{1\mu}-\frac{5}{12} \mathfrak{V}_{2\mu}-\frac{11}{6} \mathfrak{V}_{3\mu} &= P^{\nu}_{\mu} \mathcal{D}_\lambda \sigma_{\nu}{}^{\lambda}  \\
\frac{15}{9}v_{4\mu} - \frac{1}{3} v_{5\mu} -\mathfrak{V}_{1\mu}-\frac{1}{4} \mathfrak{V}_{2\mu}+\frac{1}{2} \mathfrak{V}_{3\mu} +2 R_2 &= P^{\nu}_{\mu} \mathcal{D}_\lambda \omega_{\nu}{}^{\lambda} \\
\end{split}
\end{equation}
These can be used to obtain 
\begin{equation}\label{covB:eq}
\begin{split}
\mathbf{B}^{\infty}_{\mu} &= 18 P^{\nu}_{\mu} \mathcal{D}_\lambda \sigma_{\nu}^{\lambda}+18 P^{\nu}_{\mu} \mathcal{D}_\lambda \omega_{\nu}^{\lambda}\\
\mathbf{B}^{\text{fin}}_{\mu}&= 54 P^{\nu}_{\mu} \mathcal{D}_\lambda \sigma_{\nu}^{\lambda}+90 P^{\nu}_{\mu} \mathcal{D}_\lambda \omega_{\nu}^{\lambda}\\
\mathbf{B}^{\phi}_{\mu} &= P_\mu^{\nu}u^\lambda\mathcal{D}_\nu \phi \mathcal{D}_\lambda \phi \\
\end{split}
\end{equation}
Hence, all the second-order scalar and the vector contributions to the metric can be written in terms of three Weyl-covariant scalars $ \sigma_{\alpha\beta} \sigma^{\alpha\beta},\omega_{\alpha\beta} \omega^{\alpha\beta}$ and $\mathcal{R}$ and two Weyl-covariant vectors $ \mathcal{D}_\lambda \sigma_{\mu}{}^{\lambda}$ and $\mathcal{D}_\lambda \omega_{\mu}{}^{\lambda} $ .

Further, we will enumerate below the covariant forms which appear in the tensor contributions to the metric
\begin{equation}\label{covtens:eq}
\begin{split}
\mathfrak{T}_{\mu\nu} =&\  u^{\lambda}\mathcal{D}_{\lambda} \sigma_{\mu\nu}\\
\mathfrak{T}_{5}^{\mu\nu}=&\  4(\omega^\mu{}_\lambda\omega^{\lambda\nu}+\frac{P^{\mu\nu}}{3}\omega^{\alpha\beta}\omega_{\alpha\beta})\\
\mathfrak{T}_{6\mu\nu}=&\  \sigma^\mu{}_\lambda\sigma^{\lambda\nu}-\frac{P^{\mu\nu}}{3}\sigma^{\alpha\beta}\sigma_{\alpha\beta}\\
\mathfrak{T}_{7\mu\nu}=&\ -2( \omega^{\mu}{}_\lambda\sigma^{\lambda\nu}+\omega^{\nu}{}_\lambda\sigma^{\lambda\mu})\\
(T_\phi)_{\mu\nu} &= P^\alpha_\mu P^\beta_\nu \mathcal{D}_\alpha \phi\ \mathcal{D}_\beta \phi - \frac{1}{3} P^{\alpha\beta}P_{\mu\nu} \mathcal{D}_\alpha \phi\ \mathcal{D}_\beta \phi \\
\end{split}
\end{equation}

Comparing it with the metric form in \ref{deriv:met:eq}, we find that the metric can be written in the form
\begin{equation}\label{deriv:met:eq2}
\begin{split}
ds^2 &= -2 W_1 \ u_\mu dx^\mu(dr+r \mathcal{A}_\lambda dx^\lambda) \\
     &+\left[ r^2\left( W_2 \ g_{\mu\nu} + W_3 u_\mu u_\nu  \right) + r\left( W_{4\mu} u_\nu + u_\mu W_{4\nu}\right) + W_{5\mu\nu}  \right] dx^\mu dx^\nu . \\
\end{split}
\end{equation}
where
\begin{equation}
\begin{split}
W_1 &= 1 + \frac{F_1(\xi)\sigma_{\mu\nu}\sigma^{\mu\nu}+\omega_{\mu\nu}\omega^{\mu\nu}-6\xi^2 h^{(3)}(\xi)u^\mu u^\nu \mathcal{D}_\mu\phi\ \mathcal{D}_\nu \phi}{4 r^2}+\ldots \\
W_2 &= 1 - \frac{F_1(\xi)\sigma_{\mu\nu}\sigma^{\mu\nu}+\omega_{\mu\nu}\omega^{\mu\nu}-6\xi^2 h^{(3)}(\xi)u^\mu u^\nu \mathcal{D}_\mu\phi\ \mathcal{D}_\nu \phi}{6r^2} +\ldots \\
W_3 &= \xi^{-4}+ \frac{4(F_2(\xi)-F_1(\xi)+1)  \sigma_{\mu\nu}\sigma^{\mu\nu} - 4 \omega_{\mu\nu}\omega^{\mu\nu} - \mathcal{R}}{6r^2}\\
&+\frac{ \left(\xi^2 h^{(3)}(\xi)+k^{(4)}(\xi)\right)u^\mu u^\nu \mathcal{D}_\mu\phi\ \mathcal{D}_\nu \phi+ \frac{\xi^2}{12} P^{\mu\nu} \mathcal{D}_\mu\phi\ \mathcal{D}_\nu \phi}{\xi^2 r^2}+\ldots \\
W_{4\mu} &= \frac{F_3(\xi)P_{\mu}^{\nu}\mathcal{D}_\lambda\sigma_{\nu}{}^\lambda+P_{\mu}^{\nu} \mathcal{D}_\lambda\omega_{\nu}{}^\lambda+2\xi^{-2}j^{(3)}(\xi) P_\mu^{\nu}u^\lambda\mathcal{D}_\nu \phi \mathcal{D}_\lambda \phi}{2r} +\ldots \\
W_{5\mu\nu} &= 2 r\xi F(\xi) \sigma_{\mu\nu} + W_{521}(\xi) u^\lambda\mathcal{D}_\lambda\sigma_{\mu\nu} + W_{522}(\xi) \left( \omega_{\mu}{}^{\lambda}\sigma_{\lambda\nu} +\omega_{\nu}{}^{\lambda}\sigma_{\lambda\mu} \right)\\
& + W_{523}(\xi)\left( \sigma_{\mu}{}^{\lambda}\sigma_{\lambda\nu}-\frac{P_{\mu\nu}}{3}\  \sigma_{\alpha\beta}\sigma^{\alpha\beta}\right) - \left(\omega_{\mu}{}^{\lambda}\omega_{\lambda\nu}+\frac{P_{\mu\nu}}{3}\  \omega_{\alpha\beta}\omega^{\alpha\beta}\right) \\
& + \xi^2\ln(1+\xi^{-2})\left[ C_{\mu\alpha\nu\beta}u^\alpha u^\beta + \frac{1}{4} \left(P^\alpha_\mu P^\beta_\nu  - \frac{1}{3} P^{\alpha\beta}P_{\mu\nu} \right)\mathcal{D}_\alpha \phi\ \mathcal{D}_\beta \phi \right]+ \ldots
\end{split}
\end{equation} 
with
\begin{equation}
\begin{split}
F(\xi) &= \frac{1}{4}\, \left[\ln\left(\frac{(1+\xi)^2(1+\xi^2)}{\xi^4}\right) - 2\,\arctan(\xi) +\pi\right] \\
F_1(\xi) &= 1+ 4\xi^2 \int_{\xi}^{\infty} \frac{dx}{x^5} \int_{x}^{\infty}\frac{ydy\left[1+2y(1+y+y^2)+2y^2(1+2y+3y^2)F(y)\right]}{(y+1)^2(y^2+1)^2}\\
F_2(\xi) &= 1-\frac{1}{2\xi^2}\int_{\xi}^{\infty} \frac{4(x^2+x+1)(3x^4-1)F(x)-2x^3(x^2+x+1)+x+1}{x(x+1)(x^2+1)} \\
F_3(\xi) &= 1+\frac{4}{\xi^2}\int_{\xi}^{\infty} x^3\ dx \int_{x}^{\infty}\frac{dy}{y^3(y+1)(y^2+1)}\\
\end{split}
\end{equation}
and
\begin{equation}
\begin{split}
W_{521} &= -\xi^2 \int_{\xi}^\infty\frac{dx}{x(x^4 - 1)}\int_1^x dy\ 2 y\left(\frac{1 + 3 y + 3 y^2 +3 y^3}{(y+1)(y^2+1)} - 3 y F(y)\right)\\ 
W_{522} &= -\xi^2 \int_{\xi}^\infty\frac{dx}{x(x^4 - 1)}\int_1^x dy\ 2y \left(\frac{ -1 + y + y^2 + y^3}{(y+1)(y^2+1)} - 3 y F(y)\right)\\
W_{523} &= -\xi^2\int_{\xi}^\infty\frac{dx}{x(x^4 - 1)}\int_1^x dy\ 4y \left(\frac{-1 - y + 2 y^3 + 2 y^4 + 2 y^5}{y^2(y+1)(y^2+1)} - 3 y F(y)\right)\\
\end{split}
\end{equation}
The functions $h^{(i)}(\xi),k^{(i)}(\xi)$ and $a_{(i)}(\xi)$ are as given in \eqref{duifun}.

Further, the dilaton field can be written in terms of the boundary values as
\begin{equation}
\begin{split}
\Phi& = \phi(x) + u^\mu\mathcal{D}_\mu \phi \int_{\xi}^\infty dx  \frac{x^3-1}{x^5 f(x)} 
+3 h_\phi^{(1)}(\xi)  u^\mu u^\nu \mathcal{D}_\mu \mathcal{D}_\nu \phi +\frac{3}{2} h_\phi^{(2)}(\xi) \mathcal{D}_\mu \mathcal{D}^\mu \phi\\
\end{split}
\end{equation}
where the functions $h_\phi^{(i)}(\xi)$ are as given in \eqref{duifun}.

\subsection{Constraints on the Stress Tensor and Lagrangian from Weyl Covariance}

In \eqref{stetcn} we have presented formulas for the stress tensor and the expectation
value of the dilaton upto second order in the boundary derivative expansion. In the 
notation of this section 
\begin{equation}\label{stetcnw} \begin{split}
16 \pi G_5 T^{\mu\nu}&=  (\pi \,T)^4
\left( g^{\mu \nu} +4\, u^\mu u^\nu \right) -2\, (\pi\, T )^3 \,\sigma^{\mu \nu} + 
\left( 2- \ln2 \right) (\pi T)^2 u^{\lambda}\mathcal{D}_{\lambda} \sigma^{\mu\nu}\\
& - \ln 2 (\pi T)^2  \, ( \omega^{\mu}{}_\lambda\sigma^{\lambda\nu}+\omega^{\nu}{}_\lambda\sigma^{\lambda\mu})\\
& + 2(\pi T)^2 (\sigma^\mu{}_\lambda\sigma^{\lambda\nu}-\frac{P^{\mu\nu}}{3}\sigma^{\alpha\beta}\sigma_{\alpha\beta}+C^{\mu\alpha\nu\beta}u_\alpha u_\beta) \\
& -\frac{1}{2} (\pi T)^2  (P^{\alpha\mu} P^{\beta\nu}-\frac{1}{3} P^{\alpha\beta}P^{\mu\nu}) \mathcal{D}_\alpha \phi\ \mathcal{D}_\beta \phi\\
-16 \pi G_5 e^{-\phi}{\cal L}&=(\pi T)^3\ u.\mathcal{D} \phi +  \frac{(\pi T)^2}{2}\left(\mathcal{D}^2\phi + \ln 2\ (u.\mathcal{D})^2\phi\right)
\end{split}
\end{equation}

Aspects of this result could have been predicted immediately from the requirement of 
Weyl covariance. In particular it follows immediately from the listing of 
Weyl invariants in subsection \ref{subsec:WeylTens} that at the one derivative level\footnote{
Interestingly, the fact that the Lagrangian vanishes at the zero derivative
level is not automatic from Weyl invariance, but follows instead from the 
additional dynamical input that Einstein gravity is a consistent truncation
of the Einstein-dilaton system. Once we take $\alpha'$ corrections into 
account, we do expect a contribution to the Lagrangian that is simply proportional to $T^4$.}
the stress tensor 
and the Lagrangian are proportional respectively to $(\pi\, T )^3 \,\sigma^{\mu \nu}$ 
and $(\pi T)^3 ~u.\partial \phi$ respectively. Of course the particular coefficients 
in \eqref{stetcnw} required a calculation to determine. In a similar manner the two 
derivative contributions to the stress tensor and Lagrangian are constrained by Weyl 
invariance to be linear combinations of the forms listed in \eqref{sym} and \eqref{scal}
respectively. Again the particular coefficients in \eqref{stetcnw} require knowledge
of the bulk equation of motion, and follow from explicit computation. Note in particular 
that the Lagrangian in \eqref{stetcnw}, when expressed as a linear combination of the 
terms in \eqref{scal}, turns out to have no terms proportional to the Weyl invariant scalars
$ T^{-2}\sigma_{\mu\nu}\sigma^{\mu\nu},~
T^{-2}\omega_{\mu\nu}\omega^{\mu\nu}, ~
T^{-2}\mathcal{R} ~$, 
$T^{-2}P^{\mu\nu}\mathcal{D}_\mu \phi\mathcal{D}_\nu \phi$ and $
T^{-2}u^{\mu}u^{\nu}\mathcal{D}_\mu \phi\mathcal{D}_\nu \phi$. 
Similarly, the stress tensor, when rewritten as a linear combination of the 
terms in  \eqref{sym}, has no terms proportional to any of 
$\sigma_{\mu\nu} u^\lambda\mathcal{D}_\lambda\phi$, 
$\frac{1}{2}\left[P^\alpha_\mu P^\beta_\nu + P^\alpha_\nu P^\beta_\mu - 
\frac{2}{3} P^{\alpha\beta}P_{\mu\nu} \right] \mathcal{D}_\alpha \mathcal{D}_\beta \phi$, 
or 
$\omega_{\mu}{}^{\lambda}\omega_{\lambda\nu}+\frac{P_{\mu\nu}}{3}\  \omega_{\alpha\beta}\omega^{\alpha\beta}$. Above we have already remarked that general structural features 
of the Einstein-dilaton system (e.g. the possibility of a consistent truncation 
to pure gravity) explain several of these terms from the expressions for the 
bulk metric and hence Lagrangian and the stress tensor.

\section{Causal Structure and Entropy Current}

In this brief section we will generalize the construction of 
\cite{Bhattacharyya:2008xc} to determine the location of the event horizon 
of the metric \eqref{dwitio} (under certain assumptions about the late
time behavior of the functions in this metric) and also to derive an explicit
expression for a positive divergence entropy current for the dual fluid flow. 

For the purposes of this section we restrict our attention to solutions of 
fluid dynamics for which the late time metric, dilaton, velocity and 
temperature fields all settle down to time independent constant values; we
further assume that the the event horizon of this asymptotic late time 
spacetime is given by $r=1/b(x^\mu)$. The last assumption certainly true 
for solutions that approach black branes in flat space or rotating 
black holes in global $AdS$ space at late times. Under all 
these assumptions\footnote{It is probable that these assumptions are more restrictive than necessary.It would be interesting to know which if any of these assumptions can be relaxed without affecting the simple characterization of the horizon that we give below. } the event horizon of the spacetime \eqref{dwitio} has 
a simple mathematical characterization; it is simply the unique null 
hypersurface that reduces precisely to $r=1/b(x^\mu)$ at late times 
(see \cite{Bhattacharyya:2008xc} for a related discussion). 

All the metrics we study in this paper fit into the general form given by equation 2.2 and 2.3  in 
\cite{Bhattacharyya:2008xc} with one straightforward generalization. The normalization condition $u_\mu u_\nu \eta^{\mu\nu}=-1$ of \cite{Bhattacharyya:2008xc} is replaced by $u_\mu u_\nu g^{\mu\nu}=-1$, where $g^{\mu\nu}$ is the arbitrary weakly curved boundary metric of our paper. Following \cite{Bhattacharyya:2008xc}, it is easy to work out the precise equation for the event horizon within 
the derivative expansion. In fact it turns out that equation 2.16 of \cite{Bhattacharyya:2008xc} continues to hold with the single proviso that all indices in that equation are raised by the metric 
$g^{\mu\nu}$ rather than $\eta^{\mu\nu}$. The expression 2.18 of  \cite{Bhattacharyya:2008xc}, for the normal vector to the horizon, also continues to apply with the same proviso. Finally, the area form $a$ on the event horizons in our paper may be defined exactly as in section 3.1 of 
\cite{Bhattacharyya:2008xc}. The formula  3.10 of \cite{Bhattacharyya:2008xc}
for this area form continues to apply.  Lines of constant $x^\mu$ define a map from the boundary to the horizon of our solutions. Let $s$ denote the pullback of $a$, to the boundary, under this map.
Let  $(J_s)_\mu = * s$. The following slight generalization of 3.11 of 
\cite{Bhattacharyya:2008xc} gives a formula for the entropy current 
\begin{equation}\label{eqent}
(J_s)^\mu= \frac{\sqrt{h}}{4 G_N^{(d+1)} \sqrt{g}}\frac{n^\mu}{n^v}
\end{equation}
(see  \cite{Bhattacharyya:2008xc} for the definition of $h$). 
As in \cite{Bhattacharyya:2008xc} it follows immediately from 
classic area increase theorems of event horizons in general relativity 
that $\nabla^\mu (J_s)_\mu \geq 0$. Consequently $(J_s)_\mu$ is a candidate 
entropy current with  non negative local entropy production.

We have evaluated all the abstract expressions described above for the 
specific fluid dynamical solutions presented in our paper; we present 
our results.  
In the gauge \eqref{gauge} we find the covariantization of equation 5.4 
of \cite{Bhattacharyya:2008xc}
\begin{equation}\label{horizon1}
r_h = \frac{1}{b} + \frac{b}{4}\left(S_b + \frac{1}{3}\ST5 + \frac{1}{6}\ST1^\phi\right)
\end{equation}

An explicit expression for $J^\mu_s$ is not difficult to determine: we find 
\begin{equation}\label{current}
\begin{split}
4 G b^3 J^\mu_s &= u^\mu\left(1 - \frac{b^2}{4} \left
[\left(\frac{3}{2}\ln 2 + \frac{\pi}{4}\right)^2 -1\right] \ST5  + \frac{b^2}{8}\ST1^\phi + \frac{3 b^2}{4} S_b + S_a\right)\\
&- \frac{b^2}{2} {\cal P}^{\mu\nu}\left(\nabla^\alpha\sigma_{\alpha\nu} - 3 a^\alpha\sigma_{\nu\alpha} - \frac{1}{2}{\bf B}^\phi_\nu\right) + b^2 S_c^\mu
\end{split}
\end{equation}
where
\begin{equation}\label{definition}
\begin{split}
S_b &= \frac{1}{2}{\cal S} - \frac{1}{12}\ST4 + \ST5 \left(\frac{1}{6} + {\cal C} + \frac{\pi}{6} + \frac{5 \pi^2}{48} + \frac{2}{3}\ln 2\right) + \frac{1}{12}\ST2^\phi\\
S_a &= \frac{b^2}{64}\left[-8 \ST4 - \ST5\left(8 + 48 {\cal C} + 4\pi + 4\pi^2 -36(\ln 2)^2 - 12\pi\ln2 + 16\ln 2\right)\right]\\
 &+ \frac{b^2}{8}\left(\frac{\pi}{4} + \frac{1}{2} + \ln 2\right) \ST1^\phi\\
(S_c)_\mu &= \frac{1}{16}{\bf B}^\infty_\mu - \frac{1}{144}{\bf B}^{\rm fin}_\mu -\frac{3}{8}{\bf B}^\phi_\mu
\end{split}
\end{equation}

As we have explained above, the entropy current presented in \eqref{current}
is guaranteed by the area increase theorem of general relativity to have 
non negative divergence. This claim may also be verified algebraically. 
Using the equations of motion, we find that 
\begin{equation}\label{diventcur}
\begin{split}
T \nabla_\mu J_s^\mu &=  2 \eta \left[\sigma^{\mu\nu} +\frac{\left( \pi+4+6\ \text{ln}\ 2 \right)}{16\pi\mathcal{T}} u^\lambda\mathcal{D}_\lambda \sigma^{\mu\nu} \right]^2 \\
&+\eta\left[u.\partial \phi + \frac{1}{4\pi T}(\frac{\pi}{2}+2+ 3 \ln\ 2)u^\alpha u^\beta \mathcal{D}_\alpha\mathcal{D}_\beta\phi \right]^2+\ldots \\
\end{split}
\end{equation}
where the RHS is accurate only to third order in the derivative expansion. Of course 
the RHS, which is a sum of squares,  is manifestly positive, in accordance with the 
second law. 

The second line in \eqref{diventcur} represents the production of entropy due 
to the forcing effects of a time dependent dilaton. Several comments are in order 
here. Consider the forced Navier Stokes equations $\nabla_\nu T^{\mu\nu}=f^\mu$. For an arbitrary function $f^\mu$ this equation is consistent with either local entropy increase or decrease. The fact that entropy production is always locally non negative in our situation is a consequence of the particular form of our forcing function. 

In order to see what is special about this forcing function, consider a dilaton field that is a function only of the boundary time (see $\S$\ref{sec:simp} for a more detailed discussion of this situation). According to 
\eqref{enmomforce}, the forcing function for such a field $f_\mu\propto \delta^0_\mu (\partial_0 \phi)^2$. The sign in this expression is important; it implies  that a varying dilaton 
field always pumps energy into the system. This energy raises the fluid temperature increasing its entropy.

This qualitative feature should be true of any forcing function that results from 
the coupling of a source to a gauge theory operator. A 
source term for any boundary operator excites the expectation 
value of this operator away from its thermal average. The expectation value of 
this operator then decays over a time scale $1/T$; in this process the energy 
of this excitation thermalizes, resulting in an increase in entropy.

\section{AdS$_5$ Kerr BlackHole}\label{adskerr:sec}

A three parameter set of asymptotically globally AdS$_5$ rotating black 
hole solutions has been presented in \cite{Hawking:1998kw}. These solutions are labeled by 
their temperature $T$ and their two angular velocities, $\omega_1$ and 
$\omega_2$, on $S^3$. As we will see in more detail below, these solutions 
are dual to a field theory configuration which is well described by the 
equations of fluid dynamics at large $T$ (the radius of the boundary 
sphere is unity in our conventions). 

\subsection{Summary of Results}

In this section we will rewrite the exact solutions of \cite{Hawking:1998kw}
 in the gauge 
\eqref{gauget}. The final metric admits a remarkably simple all orders 
re expression in fluid dynamical terms: it may be put in the form 
\eqref{deriv:met:eq} with 
\begin{equation} \label{bhff:eq}
\begin{split}
W_1 &= 1\ ;\qquad\ W_2 = 1+\frac{1}{3 r^2} \omega_{\alpha\beta}\omega^{\alpha\beta} \\
W_3 &= \frac{2 m}{r^4}\left(1+\frac{1}{2r^2} \omega_{\alpha\beta}\omega^{\alpha\beta}\right)^{-1} -\frac{2}{3 r^2} \omega_{\alpha\beta}\omega^{\alpha\beta}-\frac{\mathcal{R}}{6 r^2} \\
W_{4\mu} &= -\frac{1}{2r}P^\nu_\mu\mathcal{D}_\lambda \omega^\lambda{}_\nu \\
W_{5\mu\nu} &= -\left(\omega_\mu{}^\lambda\omega_{\lambda\nu} +\frac{\omega_{\alpha\beta}\omega^{\alpha\beta}}{3} P_{\mu\nu} \right)
\end{split}
\end{equation}
Moreover the radial location event horizon of this metric is 
given by the solution to the equation 
\begin{equation}\label{loceh}
(1+\frac{1}{r_H^2})(r_H^2+ \omega_1^2)(r_H^2+\omega_2^2)
=2m
\end{equation}
The entropy of this black hole is given by 
\begin{equation}\label{entbh} 
S=\frac{2 m \Omega_3}{r_H (1-\omega_1^2)(1-\omega_2^2) (1+\frac{1}{r_H^2})}
\end{equation}
Moreover the various components of the local entropy current defined in 
\cite{Bhattacharyya:2008xc} evaluate, on this rotating black hole solution 
\begin{equation}\label{exentcur}\begin{split}
J^t&=\frac{(r_H^2+\omega_1^2)(r_H^2+\omega_2^2)}{r_H} \\
J^\phi&=\frac{\omega_1 (r_H^2+\omega_1^2)(r_H^2+\omega_2^2) (1-\omega_1^2) 
}{r_H(r_H^2+\omega_1^2)} \\
J^\psi&=\frac{\omega_2 (r_H^2+\omega_1^2)(r_H^2+\omega_2^2)(1-\omega_2^2)}
{r_H (r_H^2 +\omega_2^2)}\\
\end{split}
\end{equation}

Notice that the exact result \eqref{bhff:eq} admits a remarkably simple 
derivative expansion. The derivative expansion of the functions 
$W_1$, $W_2$, $W_{4\mu}$ and $W_{5\mu\nu}$ terminates at second order in 
derivatives. On the other hand the function $W_3$ includes terms at all orders
in this expansion: however this expansion sums up to a simple geometrical 
series. Several additional remarks are in order 
\begin{enumerate}
\item We check below that  the expansion of \eqref{bhff:eq} to 
second order in the derivative expansion matches perfectly with the general 
predictions given in \eqref{deriv:met:eq}. This is a nontrivial 
consistency check on the main result of this paper. 
\item Note that \eqref{bhff:eq} receives contributions from terms 
that depend on the curvature of the boundary metric only at second order in 
the derivative expansion. Consequently the second order fluid results, 
\eqref{bhff:eq} listed above, exactly account for all curvature dependencies 
in the exact result \eqref{deriv:met:eq}. In particular the $AdS$ Schwarzschild 
solution is reproduced exactly at second order in the fluid expansion.  
\item The fluid dynamical expansion of the metric \eqref{bhff:eq} 
is convergent for $2 r^2 \geq \omega_{\alpha \beta}\omega^{\alpha \beta}$. 
It follows that this expansion is convergent everywhere outside the 
horizon for $T^2 \gg \omega^{\alpha \beta}\omega_{\alpha \beta}$. 
\end{enumerate}
Expressions for the location of the event horizon and the local 
entropy current may also be expanded rather simply, to all orders, in 
fluid dynamical expansion. We demonstrate below that the 
expansion of these expressions to second order also agree with the 
predictions of the previous section.

In summary the fluid expansion works even better for the exact solution 
\eqref{deriv:met:eq} than we might naively have the right to expect. 
In the rest of this section we will derive and explain all these results 
in detail.

\subsection{Rotating Black Holes in Fluid Dynamical Coordinates}
\subsubsection{The AdS Kerr Solution  and its Stress tensor}
The asymptotically globally AdS$_5$ rotating black hole solution presented 
in \cite{Hawking:1998kw} is given by 
\begin{equation}\label{ads5kerr:eq}
\begin{split}
ds^2 &= -\frac{\Delta_r}{\rho^2}\left( d\hat{t}-\frac{\omega_1 \sin^2\Theta}{1-\omega_1^2}d\hat{\phi} -\frac{\omega_2 \cos^2\Theta}{1-\omega_2^2}d\hat{\psi} \right)^2 
+ \frac{\rho^2}{\Delta_r} dr^2 + \frac{\rho^2}{\Delta_\Theta} d\Theta^2 \\
&+ \frac{\Delta_\Theta}{\rho^2}\left( \sin^2\Theta \left(\omega_1 d\hat{t}-\frac{r^2+ \omega_1^2}{1-\omega_1^2}d\hat{\phi}\right)^2 + \cos^2\Theta \left(\omega_2 d\hat{t}-\frac{r^2+ \omega_2^2}{1-\omega_2^2}d\hat{\psi}\right)^2 \right) \\
&+ \frac{1+r^2}{r^2\rho^2}\left( \omega_2 \sin^2\Theta \left(\omega_1 d\hat{t}-\frac{r^2+ \omega_1^2}{1-\omega_1^2}d\hat{\phi}\right) + \omega_1 \cos^2\Theta \left(\omega_2 d\hat{t}-\frac{r^2+ \omega_2^2}{1-\omega_2^2}d\hat{\psi}\right) \right)^2
\end{split}
\end{equation}
where
\begin{equation}\label{kerrparam:eq}
\begin{split}
\rho^2 &\equiv r^2 + \omega_1^2 \cos^2\Theta +\omega_2^2 \sin^2\Theta\\
\Delta_r &\equiv \frac{1}{r^2}(1+r^2)(r^2 + \omega_1^2)(r^2 + \omega_2^2)- 2m\\
\Delta_\Theta &\equiv 1- \omega_1^2 \cos^2\Theta -\omega_2^2 \sin^2\Theta\\ 
\end{split}
\end{equation}

In the large $r$ limit (i.e. near the boundary) the induced metric on
a surface of constant $r$ is given by 
\begin{equation}\label{ads5kerrbnd:eq}
\begin{split}
\frac{ ds^2_{\text{Bnd}}}{r^2} &= g_{\mu\nu} dx^\mu dx^\nu\\
&= -\frac{\Delta_\Theta d\hat{t}^2}{(1-\omega_1^2)(1-\omega_2^2)}
+ \frac{d\Theta^2}{\Delta_\Theta} 
+ \frac{\sin^2\Theta}{1-\omega_1^2}(d\hat{\phi}+\omega_1 d\hat{t})^2 
+ \frac{\cos^2\Theta}{1-\omega_2^2}(d\hat{\psi}+\omega_2 d\hat{t})^2 \\
\end{split}
\end{equation}
Consequently the solution \eqref{ads5kerr:eq} is dual to a state of 
the CFT on the space Weyl equivalent to \eqref{ads5kerrbnd:eq}. 
\eqref{ads5kerrbnd:eq} describes the so called Rotating Einstein Universe.
 
The boundary stress tensor \eqref{stetc} dual to this metric is easily evaluated, and may be 
written in the form  
\begin{equation} \label{stu}
\begin{split}
T_{\mu\nu} &= \frac{m}{8\pi G_5} (g_{\mu\nu} + 4 u_\mu u_\nu) +\frac{1}{64\pi G_5}\left(R_{\alpha\beta}R^{\alpha\mu\beta\nu}-\frac{R^2}{12} g_{\mu\nu}\right) \qquad \text{where} \\
-u_\mu dx^\mu &= d\hat{t}-\frac{\omega_1 \sin^2\Theta}{1-\omega_1^2}d\hat{\phi} 
-\frac{\omega_2 \cos^2\Theta}{1-\omega_2^2}d\hat{\psi}\\
&= \frac{\Delta_\Theta d\hat{t}}{(1-\omega_1^2)(1-\omega_2^2)}-\frac{\omega_1 \sin^2\Theta}{1-\omega_1^2}(d\hat{\phi}+\omega_1 d\hat{t}) 
-\frac{\omega_2 \cos^2\Theta}{1-\omega_2^2}(d\hat{\psi}+\omega_2 d\hat{t}) \\
u^\mu \partial_\mu &= \frac{\partial}{\partial\hat{t}}
\end{split}
\end{equation}
where $R_{\mu\nu\lambda\sigma}$ denotes the boundary curvature tensor associated with the metric $g_{\mu\nu}$. Apart from an additive curvature dependent piece whose form is dictated by 
the conformal anomaly\footnote{Note, for example, that the trace of the stress tensor in this background as calculated from above\cite{Henningson:1998gx,Awad:2000aj} is $ 64\pi G_5 T^\mu_\mu = R_{\alpha\beta}R^{\alpha\beta}- \frac{R^2}{3}$.}, this stress tensor is precisely that for a perfect 
conformal fluid with velocity vector $u^\mu=(1,0,0,0)$ and pressure 
$p=\frac{m}{8\pi G_5}$.

\subsubsection{Relationship to analysis in \cite{Bhattacharyya:2007vs}}

We pause here to connect these results to those derived in 
\cite{Bhattacharyya:2007vs}. The authors of that paper analyzed the same 
rotating black hole solutions, however they worked with a different set of 
coordinates denoted here by ($\tilde{r}, \hat{t}, \phi', \psi', \Theta'$).  In these 
coordinates the boundary of the rotating black hole solutions is naturally $S^3 \times R$ 
as we now explain.

The Rotating Einstein Universe is Weyl Equivalent to $S^3 \times R$. In order to see this 
consider the boundary coordinate transformation 
\begin{equation}\label{bct} \begin{split}
\cos^2 \Theta&=\frac{(1-\omega_2^2)\cos^2 {\Theta'}}{1-\omega_1^2 
\sin^2 {\Theta'} -\omega_2^2 \cos^2 {\Theta'} }  \\
\sin^2 \Theta&=\frac{(1-\omega_1^2)\sin^2 {\Theta'}} {1-\omega_1^2 
\sin^2 {\Theta'} -\omega_2^2 \cos^2 {\Theta'} }\\
\hat{\phi}+\omega_1 \hat{t}&={\phi'}\\
\hat{\psi}+\omega_2 \hat{t}&={\psi'}\\
\end{split}
\end{equation}
Expressed in terms of these new variables, the boundary metric 
\eqref{ads5kerrbnd:eq} may be written as
\begin{equation}\label{nbm1} \begin{split}
\frac{ ds^2_{\text{Bnd}}}{r^2}
&=\frac{1}{1-\omega_1^2 \sin^2 {\Theta'} -
\omega_2^2 \cos^2 {\Theta'} }
\left(- d\hat{t}^2 +  d {\Theta'}^2 + \cos^2 {\Theta'} 
d {\psi'}^2
+ \sin^2 {\Theta'} d {\phi'}^2 \right)\\
&=\gamma^2\left( -d\hat{t}^2 + d \Omega_3^2 \right)  \\
\gamma&=\frac{1}{\sqrt{1-\omega_1^2 \sin^2 {\Theta'} -
\omega_2^2 \cos^2 {\Theta'} }}
\end{split}
\end{equation}
where $d\Omega_3^2$ represents the usual metric on the round 3 sphere. 

In order to obtain a boundary metric that is actually that on $S^3 \times R$ rather 
than simply conformal to it, the authors of \cite{Bhattacharyya:2007vs} worked with the 
redefined radial variable 
${\tilde r} =\frac{r}{\sqrt{1-\omega_1^2 \sin^2
{\tilde  \Theta} -\omega_2^2 \cos^2 
{\Theta'} }} =\frac{r}{\sqrt{1-\omega_1^2 \cos^2
{ \Theta} -\omega_2^2 \sin^2
{\Theta} }}$ so that the induced metric on slices of large constant ${\tilde r}$ is 
equal to  
\begin{equation} \label{nbm2} 
\frac{ds^2}{{\tilde r}^2}  =-dt^2 +d\Omega_3^2.
\end{equation}

With these conventions the rotating black hole solution is dual to a fluid flow 
on $S^3 \times $ time. The fluid velocities and temperatures of this flow are 
given by acting upon the velocity and temperature fields $u^\mu=(1,0,0,0)$ $T=\frac{(2M)^{\frac{1}{4}}}{\pi}$, of the previous 
subsection,  with the coordinate transformation $\eqref{bct}$ 
followed by a Weyl transformation (compare \eqref{nbm1} and \eqref{nbm2}) 
yielding 
\begin{equation}\label{newvel} \begin{split}
{\tilde u}&= \gamma\left(\frac{\partial}{\partial \hat{t}} 
+\omega_1\frac{\partial}{\partial \phi'} + 
\omega_2 \frac{\partial}{\partial \psi'} \right) \cr
{\tilde T}& = \frac{(2M)^{\frac{1}
{4} }}{\pi} \gamma \cr
\gamma & =\frac{1}{\sqrt{1-\omega_2^2 \cos^2
{\tilde  \Theta} -\omega_1^2 \sin^2 {\Theta'} }}
\end{split}
\end{equation}
precisely as reported in \cite{Bhattacharyya:2007vs}.

\subsubsection{Recasting the solution in the Fluid mechanical gauge}
Continuing with our analysis of \eqref{ads5kerrbnd:eq}, we will now 
recast this metric in the gauge \eqref{gauget}. As was explained in \cite{Bhattacharyya:2008xc}, the 
coordinates in gauge \eqref{gauget} are adapted to a congruence of null 
ingoing geodesics, whose tangent vectors near the boundary are given by 
$\frac{dz}{d\lambda}=1$, $\frac{dx^\mu}{d\lambda}=u^\mu$ in Graham Fefferman
coordinates. While it turns out to be difficult to solve for the most general 
null geodesic in \eqref{ads5kerrbnd:eq}, precisely this congruence turns out 
to be easy to determine and is given by 
\begin{equation}\label{geodesics} \begin{split}
\frac{dr}{d \lambda}&=-1 \\
\frac{d \Theta}{d \lambda}&=0 \\
\frac {dt}{d r} & = - \frac{(r^2+\omega_1^2)(r^2+\omega_2^2)}{r^2 \Delta_r} \\
\frac {d\phi}{dr}&=\omega_1 \frac{ (1-\omega_1^2)}{r^2+\omega_1^2} 
\frac {dt}{d r}\\
\frac {d\psi}{dr}&=\omega_2 \frac{ (1-\omega_2^2)}{r^2+\omega_2^2} 
\frac {dt}{d r}\\
\end{split}
\end{equation}
As we have explained before, in the gauge \eqref{gauget} the coordinates 
$x^\mu$ are constant along this congruence of geodesics, and the coordinate
$r$ is simply the affine parameter along these geodesics. Consequently, 
it follows that the change of variables to the 
Eddington-Finkelstein like co-ordinates \eqref{gauget} is given by  
\begin{equation}\label{KerrEddFink:eq}
\begin{split}
d\hat{t} &= dt - \frac{(r^2 + \omega_1^2)(r^2 + \omega_2^2)}{r^2 \Delta_r} dr \\
d\hat{\phi} &= d\phi - \frac{(1- \omega_1^2)(r^2 + \omega_2^2)}{r^2 \Delta_r} dr\\
d\hat{\psi} &= d\psi - \frac{(r^2 + \omega_1^2)(1- \omega_2^2)}{r^2 \Delta_r} dr\\
\end{split}
\end{equation}
Expressed in terms of the new coordinates the metric becomes
\begin{equation}\label{ads5kerrEF:eq}
\begin{split}
ds^2 &= 2dr\left( dt-\frac{\omega_1 \sin^2\Theta}{1-\omega_1^2}d\phi 
-\frac{\omega_2 \cos^2\Theta}{1-\omega_2^2}d\psi \right)\\
&-\frac{\Delta_r}{\rho^2}\left( dt-\frac{\omega_1 \sin^2\Theta}{1-\omega_1^2}d\phi -\frac{\omega_2 \cos^2\Theta}{1-\omega_2^2}d\psi \right)^2 + \frac{\rho^2}{\Delta_\Theta} d\Theta^2 \\
&+ \frac{\Delta_\Theta}{\rho^2}\left( \sin^2\Theta \left(\omega_1 dt-\frac{r^2+ \omega_1^2}{1-\omega_1^2}d\phi\right)^2 + \cos^2\Theta \left(\omega_2 dt-\frac{r^2+ \omega_2^2}{1-\omega_2^2}d\psi\right)^2 \right) \\
&+ \frac{1+r^2}{r^2\rho^2}\left( \omega_2 \sin^2\Theta \left(\omega_1 dt-\frac{r^2+ \omega_1^2}{1-\omega_1^2}d\phi\right) + \omega_1 \cos^2\Theta \left(\omega_2 dt-\frac{r^2+ \omega_2^2}{1-\omega_2^2}d\psi\right) \right)^2
\end{split}
\end{equation}

As we describe in detail in the appendix \ref{app:adskerr}, we conclude that the 
AdS$_5$-Kerr metric can be written in the following manifestly Weyl-covariant form
\begin{equation} \label{bhfr:eq}
\begin{split}
ds^2 &= -2 u_\mu dx^\mu (dr+r \mathcal{A}_\lambda dx^\lambda) + r^2 g_{\mu\nu} dx^\mu dx^\nu\\
&-\left(u_\mu \mathcal{D}_\lambda\omega^\lambda{}_\nu + \omega_\mu{}^\lambda\omega_{\lambda\nu} + \frac{\mathcal{R}}{6} u_\mu u_\nu \right)dx^\mu dx^\nu + \frac{2 m}{r^2}\left(1+\frac{1}{2r^2} \omega_{\alpha\beta}\omega^{\alpha\beta}\right)^{-1} u_\mu u_\nu dx^\mu dx^\nu \\
&=-2 W_1 u_\mu dx^\mu(dr+r \mathcal{A}_\lambda dx^\lambda) \\
&+\left[ r^2\left( W_2 \ g_{\mu\nu} + W_3 u_\mu u_\nu  \right) + r\left( W_{4\mu} u_\nu + u_\mu W_{4\nu}\right) + W_{5\mu\nu}  \right] dx^\mu dx^\nu . \\
\end{split}
\end{equation}
with the functions in these equations given by \eqref{bhff:eq}.

\subsection{Horizon and Entropy Current}

As we have mentioned above, the location of the event horizon (in the radial 
variable of the gauge \eqref{gauget}) is given by the solution to the 
equation \eqref{loceh}. The solution to this equation is easily obtained 
at second order in the derivative expansion: we find 
\begin{equation}\label{ehso} \begin{split}
r_H& = \frac{1}{b}\left( 1-\frac{b^2}{4}(1+\omega_1^2+\omega_2^2) \right) \\
&= \frac{1}{b}\left(1-\frac{b^2}{4}\left( \frac{3}{2} 
\omega_{\alpha \beta} \omega^{\alpha \beta} + \frac{\mathcal{R}}{6} \right)
\right)\\
\frac{1}{b}&=(2 m)^{\frac{1}{4}}\\
\end{split}
\end{equation}
It is easy to check that this result agrees with our general prediction 
\eqref{horizon1}, once that equation is re-expressed in the radial variable 
of the  \eqref{gauget}
\begin{equation}\label{horizon2}
r_h = \frac{1}{b} + 
\frac{b}{4}\left(S_b + \frac{1}{3}\ST5 + \frac{1}{6}\ST1^\phi\right)
 + \frac{3 b^2}{2} \int_{r=1/b}^\infty H(br)
\end{equation}
On the velocity configuration dual to the black hole $\mathfrak{S}_5$ 
and  $\mathfrak{S}_5$ vanish and 
$$S_b=-\frac{ \omega_{\alpha \beta}\omega^{\alpha\beta}}{2}
-\frac{\mathcal{R}}{6}, ~~~\int_{r}^\infty 
H(r)=-\frac{\omega_{\alpha\beta}\omega^{\alpha \beta}}{4r}; $$
inputting these values we see that \eqref{horizon2} and \eqref{ehso} 
agree.

In a similar fashion, the formula for the entropy and entropy 
current \eqref{entbh} for these 
black holes may be expanded to second order: we find 
\begin{equation}\label{entso} \begin{split}
S&=\frac{\Omega_3}{b^3(1-\omega_1^2)(1-\omega_2^2)} 
\left( 1+ \frac{b^2}{4} \left( \frac{3 \omega_{\alpha \beta}
\omega^{\alpha \beta}}{2} + \frac{\mathcal{R}}{6} -4 \right)\right)\\
J^t&=\frac{1}{b^3} \left( 1+\frac{b^2}{4}(\omega_1^2+\omega_2^2-3) \right)\\
J^\phi& =\frac{\omega_1 (1-\omega_1^2)}{b}\\
J^\psi& =\frac{\omega_2 (1-\omega_2^2)}{b}
\end{split}
\end{equation}
On the other hand the general formula for the entropy current 
\eqref{current} evaluates, on the specific fluid flow at hand, to 
\begin{equation}\label{ecso}
4 G_5 b^3 J_s^\mu = u^\mu \left( 1- \frac{b^2}{8}\left(5 \omega_{\alpha \beta}
\omega^{\alpha \beta} + \mathcal{R}\right) \right) 
+ \frac{b^2}{2} P^{\mu \nu} \mathcal{D}_\lambda \omega_\nu{}^\lambda
\end{equation}
It is easy to verify, using \eqref{weylREU:eq}, that \eqref{ecso} and 
\eqref{entso} agree. Upon integration $S=\int \sqrt{g} J^0$ with 
$\sqrt{g}= \frac{\sin \Theta \cos \Theta}{(1-\omega_1^2)(1-\omega_2^2)}$
we also reproduce the first of  \eqref{entso}.

\section{Some Simple Solutions of Forced Fluid Dynamics}\label{sec:simp}

In this section we construct some simple solutions to the equations of motion of fluid 
mechanics with forcing terms derived above, and consider their interpretation in the bulk. 
In order to make the physical points of interest most immediately, in this 
section, we often  work with the crudest approximations that capture the physics at hand. 
We postpone a more careful study of these solutions (and hopefully several others)
to future work. 

In subsections \ref{subs:metforcing}  we study fluid flows that are pushed to high Reynolds numbers 
by the effective forcing effect of a varying background metric. In subsection \ref{sub:timedepdil} we 
study `cosmological' solutions of fluid dynamics corresponding to a time dependent 
but spatially homogeneous dilaton. As we have explained in the introduction, we believe 
that these solutions qualitatively capture the excursion of the bulk geometry into 
regions of strong curvature. Finally, in subsection \ref{subs:spacevardil} we study a fluid that is pushed 
into motion by the forcing effect of a varying dilaton field. 

\subsection{Metric as the Forcing Term}\label{subs:metforcing}

\subsubsection{Hydrostatic Solution in an arbitrary spacetime}

We begin by considering the case where the forcing term arises due to a non-trivial metric. 
We start with the stress tensor in the perfect fluid approximation,
\be
\label{pfs}
T^{\mu\nu}=(\pi T)^4(g^{\mu\nu}+4u^\mu u^\nu),
\ee
in the presence of a time independent metric of the form,
\be
\label{tindme}
ds^2=g_{00}dt^2+g_{ij}dx^idx^j.
\ee
It is easy to see that the conservation equations,
\be
\label{conse}
\nabla_\mu T^{\mu\nu}=0,
\ee
 then admit a hydrostatic solution, with  four-velocity,
\be
u^0=\sqrt{-g^{00}}, u^i=0,
\ee
and  temperature,
\be
\label{temss}
T={\frac{C}{\sqrt{|g_{00}|}}}.
\ee
The temperature dependence can be understood as arising   due to the gravitational red-shift.
We see that the temperature is higher,   where $|g_{00}|$ is smaller, i.e., where
 the gravitational potential is deeper.
In the bulk the radial location of the horizon is given by,
$r=\pi T$, and becomes a function of the spatial coordinates.

This solution gets corrections due to the additional terms in the stress tensor,
 eq.(\ref{stetc}). Since the fluid is at rest the
viscosity term is irrelevant. Although we have not attempted this, it 
 should be possible to work out the corrections due to  the  second-order 
terms found in this paper in  a straightforward manner. All such terms 
are suppressed by a factor of $(TR)^{-2}$, where $R$ is the length scale 
associated with the boundary spacetime, compared to the leading order result 
presented above. These terms will lead to a corresponding second-order 
correction to the dual geometry.

\subsubsection{Small deviations from hydrostatic equilibrium}

We now specialize to a boundary  metric of the form
\begin{equation}
\label{spmet} \begin{split}
g_{ij}&=\delta_{ij} \\
g_{00}&=g_{00}(z)\\
\end{split}
\end{equation}
($z$ is one of the spatial coordinates) and study a particular 
perturbation about hydrostatic equilibrium. In particular, we search 
for a steady state solution in which, in addition to a varying temperature, 
the fluid has a small, time independent velocity purely in the $z$ direction. 
Throughout this subsubsection we work to first order in this spatial 
velocity $u^z$. 

The $\nu=0$ component of eq.(\ref{conse}) gives \footnote{We work in the perfect fluid 
approximation and at first order in velocities. It should be straightforward to 
account for the first nonzero corrections to these approximations.}
\be
\label{velsol}
T^4 u^z |g_{00}|=\tilde{c},
\ee
where $\tilde{c}$ is an integration constant. 
The $\nu=z$ component of eq.(\ref{conse})
 is unchanged from the hydrostatic case, at linear order, and gives, again, 
 eq.(\ref{temss}). From, eq. (\ref{velsol}), (\ref{temss}) it follows that  
\be
\label{detvz}
u^z=u^z_0 |g_{00}|,
\ee
where $u^z_0$  is a constant. (\ref{temss}), and (\ref{detvz}), determine
the temperature and velocity  $T, u^z,$ as a function of $z$. 
Let us suppose that $g_{00}=-1$ at a particular location and 
changes to $-(1+f)$ over the length scale $L$. It follows that the change 
in the fluid velocity over the same length scale equal in magnitude 
to $f u^z_0$. While $u^z_0$ has been assumed to be small 
compared to unity in this subsection, and while we might wish to restrict
$f$ also to be small compared to unity for some physical purposes, 
it is consistent to hold each of these quantities fixed as $TL$ is taken 
large. In this case the formal `Reynolds number' of this flow  
\be
\label{reynumber}
R_e\sim \Delta v TL,
\ee
becomes parametrically large in the fluid dynamical limit $TL \gg 1$.

While the static flow described above is rather tame, we believe it illustrates the 
general point argued for in the introduction, namely that even though the forcing 
effects of gravity are mild (suppressed by $1/L$) they can build up to ${\cal O}(1)$
changes in the velocity  - and hence high Reynolds numbers - 
over length scale $L$. Consequently one might hope to be able to stir the fluid 
into steady state within the validity regime of the approximations of this paper, 
though we would probably need a more general metric (one that depends on several 
spatial directions as well as time) for this purpose.

\subsubsection{Flows forced by a time dependent spatial metric}

We can also consider an analogous situation where the metric is,
\be
\label{anmet}
ds^2=-dt^2+g_{zz}(t)dz^2+dx^2+dy^2,
\ee
with only one non-trivial spatial component that depends on time. A consistent solution 
can be obtained with $u^0,u^z$ being the only non-zero components of the $4$-velocity. 
Working with the perfect fluid stress tensor, one finds from the $\nu=0$ component of 
eq.(\ref{conse})
that, 
\be
\label{int1}
T^4u^0u^z =\frac{c_1}{ (g_{zz})^{3/2}}.
\ee
The four-velocity satisfies the constraint, 
\be
\label{fvelc}
(u^0)^2-(u^z)^2 g_{zz}=1,
\ee
The last  equation is the  $\nu=z$ component of eq.(\ref{conse}).
Using eq.(\ref{int1}), eq.(\ref{fvelc}), to eliminate two of the variables, one can reduce this
problem to quadrature. We skip some of the details here. 

The temperature and $4$-velocity 
are time dependent in this case. It is easy to see that  
again order unity changes in the 
metric, $g_{zz}$, result in order unity changes in the velocity. The time varying metric
allows energy to be pumped in or out of the system. If the time scale over which this happens
is big compared to the temperature, as is needed for the fluid mechanics approximation to be 
valid, one finds analogous to the the spatially varying case discussed above, that the 
Reynolds number is  much bigger than unity.

\subsection{ Time Dependent Dilaton and Highly Curved Spacetimes}\label{sub:timedepdil}

We now study a simple `cosmological' solution of fluid dynamics. In the solution 
we study the metric is taken simply to be $\eta_{\mu\nu}$, but the dilaton 
is taken to be a slowly varying function of time. We are interested in answering 
the following question: suppose we start with field theory heated up to
a temperature $T$ and at coupling constant $\lambda_1$. Suppose we then slowly 
vary the coupling from $\lambda_1$ to $\lambda_2$ over a time $\Delta t$. What 
is the final state of the theory at the end of this process?

Using the equations of fluid dynamics derived in this paper, it is easy to 
answer this question (we comment below on the validity of these equations for the 
purposes of this issue). In particular, there exists a solution to the fluid 
equations of motion in which the fluid is always at rest and the temperature remains 
spatially homogeneous but heats up slowly in time, in response to the varying dilaton. 
In more detail consider the configuration 
\be
\label{fve}
u^0=1, \ u^i=0.
\ee
Taking the leading contribution to the dilaton forcing function (this is the first 
term on RHS of the second equation in eq.(\ref{stetc})) 
gives, from eq.(\ref{enmomforce}),
\be
\label{dilffl}
\nabla_\mu T^{\mu\nu}=-(\pi T)^3 \partial^\nu\phi \partial_0\phi.
\ee
If we take the perfect fluid stress tensor, eq.(\ref{pfs}), and $4$-velocity, eq.(\ref{fve}), 
this gives rise to one non-trivial equation for $T$,  which can be integrated to give,
\be
\label{eqt}
T(t_f)-T(t_i)=\frac{1}{ 12 \pi} \int_{t_i}^{t_f} (\dot{\phi})^2 dt.
\ee

We see that, irrespective of the details, 
 time dependence of the dilaton always increases the temperature and thus the 
energy and the entropy density of the fluid monotonically.  In the bulk correspondingly 
the horizon area  increases monotonically. This ties in with the discussion below 
\eqref{diventcur}. 

Note that 
\be
\label{smc}
\frac{\Delta T}{T}\sim \frac{(\Delta \phi)^2 }{ T \Delta  t} \ll 1.
\ee
i.e. a total change in the dilaton, $\Delta \phi$, over a time, $\Delta t$, results in a  
small fractional change in temperature if $\delta t$ is sufficiently large. When the 
LHS of \eqref{smc} is small the condition for the validity of fluid dynamics 
\be
\label{fvalid}
\frac{\dot{T}}{T^2}= \frac{(\Delta \phi)^2}{(T \Delta t)^2} \ll 1.
\ee
is automatically met provided $\Delta \phi \ge 1$ and hence 
$T \Delta t>1$. 

In summary a time dependent coupling constant always heats up the system; however the 
fractional increase in the temperature can be arranged to be small - even for a large 
fractional increase in the coupling - provided the coupling constant is changed relatively 
slowly.

Corrections to eq.(\ref{eqt}) will arise because of corrections in the stress tensor, eq.(\ref{stetc}),
and in the forcing function, eq.(\ref{enmomforce}). If the condition, eq.(\ref{fvalid}), is valid 
corrections due to second order terms  in the stress tensor are small \footnote{Any corrections due
to the  viscosity term vanish in this case.}.
Corrections in the forcing function go like, $T^2 \dot{\phi} \ddot{\phi}$. These are small,
compared to the leading order term,  if $T \Delta t>1$.

\subsubsection{Highly Curved Spacetimes}

In the bulk of this paper we have derived a set of fluid dynamical equations from the Einstein Hilbert 
Lagrangian and studied some aspects of the dynamics of these equations. Our analysis so far has been 
quantitative. On the other hand, in this subsubsection we attempt to  explore some 
qualitative aspects of bulk dynamics of highly curved spacetimes via fluid dynamics.

In particular, we wish to investigate the behavior of a set of classical solutions of string theory (rather than gravity) that is dual to evolutions of large $N_c$ ${\cal N}=4$ Yang Mills theory with a time varying 'tHooft coupling that is lowered down to order unity. The bulk equations that govern these systems are unknown. Nonetheless we will attempt to qualitatively understand these evolutions  using an expectation based on physical intuition : at long wavelengths these (unknown) equations of classical string theory should reduce to the equations of boundary fluid dynamics with $\lambda$ dependent parameters. Though we do not know the detailed $\lambda$ dependence of these parameters,  we will attempt to estimate their qualitative properties to the extent needed for our analysis. 

We pause here to explain this intuition in more detail. The equations of fluid dynamics follow simply from 
symmetries combined with the physical expectation of local equilibration. Any interacting system is 
expected to equilibrate locally over a length and time scale set by its `mean free path' which may be crudely estimated by the ratio $\eta/\rho$ where $\eta$ is the viscosity and $\rho$ the energy density of the system. It follows from  'tHooft scaling and dimensional analysis that this ratio takes the form 
$q(\lambda)/T$ for some function $q(\lambda)$ in ${\cal N}$=4 Yang Mills theory. We know from perturbation theory that $q(\lambda)$ diverges at weak coupling, and we know  from the AdS/CFT correspondence that $q(\lambda)$ is a constant at strong coupling. From these two behaviors, it seems reasonable to guess that $q(\lambda)$ decreases monotonically as the coupling is increased or,  at least, that it never diverges at a finite value of $\lambda$. If this last guess is correct, the effective equilibration length scale for ${\cal N}=4$ Yang Mills should be of order $1/T$ for all $\lambda$ greater than or of order unity, and  fluid dynamics should be a good description of any evolution of the system in which all quantities vary on a length scale that is large compared to $1/T$.We make this assumption in what follows below. We also assume the absence of any phase transition in high temperature ${\cal N}=4$ Yang Mills as $\lambda$ is varied from large values to values of order unity. We wish to emphasize that if either of the assumptions listed in this paragraph fails, none of the conclusions we reach in this section need apply.

With all these caveats in mind, we now proceed to consider a situation where we start,
in the far past, with a value of the dilaton on the boundary such that $gs N \gg 1$.
We take the dilaton to be  a slowly varying function of time
and reduce its value till,  $g_s N = \lambda_{min}$. Then  increase it again  so that in the 
far future $g_s N \gg 1$.

The fluid dynamical solution presented above accurately captures the boundary dynamics of 
this situation provided $\lambda_{min} \gg 1$. When this condition is met, the 
results of this paper eq.(\ref{dwitio}) also yield the bulk dual of this field theoretic 
evolution. 

Now let us instead consider take $\lambda_{min}$ to be ${\cal O}(1)$. In this case 
we expect the evolution described in this subsection to be given by some classical 
solution of string theory rather than supergravity. We also expect spacetime curvatures
to become of order string scale - and so for stringy effects to become very important - 
over the times at which $\lambda$ is of order $\lambda_{min}$. However curvatures are 
small in string units (and so the SUGRA approximation is good) at early and late times. 
In other words the dual bulk solution is expected to be a cosmology where
an initially smooth space-time becomes highly curved, with a curvature of order the 
string scale so that it is not well described by the two derivative
SUGRA approximation, and then returns in the future to being smooth again. 
\footnote{For some recent work
on time dependent cosmologies in the context of AdS/CFT, see \cite{Ishino:2005ru,Craps:2005wd,Das:2006dz,Chu:2006pa,Das:2006pw,Craps:2006xq,Ishino:2006nx,Kodama:2006bw,Chu:2007um,Turok:2007ry,Awad:2007fj,Craps:2007ch}.}
We now want to inquire what information we can glean about this evolution process 
by the use of boundary fluid dynamics. 

Of course the specific fluid dynamical equations we have derived in this paper are only valid 
at very large $\lambda$. However aspects of these equations are dictated merely by 
symmetries and the physical expectations of equilibration, and may have 
much greater validity. Of special importance to us in this section is the structure of 
dilaton dependence in the forcing term in the Navier Stokes $\nabla_\mu T^{\mu\nu}=f^\nu$ 
studied in this paper. As we have explained in the introduction, it follows from general 
field theoretic reasoning that $f^\mu= e^{-\phi}{\cal L}\partial^\mu \phi $ where ${\cal L}$ is 
the expectation value of the Lagrangian. As we have explained above, it 
further follows Weyl invariance that upto the one derivative level 
$e^{-\phi}{\cal L}= T^4 r(\lambda)+ T^3 g(\lambda) u. \partial \phi$. Similar statements apply to the form of the 
stress tensor upto first order in the derivative expansion. 
It follows that the leading two (boundary) derivative equation of motion that 
governs the evolution described in this subsection is 
\be
\label{dilfflex}
\frac{1}{T}\frac{\partial T}{\partial t}=-h(\lambda) \frac{(\partial_0\phi)^2}{12\pi T}
+g(\lambda) \partial_0\phi .
\ee
where $h(\lambda)$ is an unknown function that evaluates to unity at $\lambda = \infty$ 
and is expected to be of order unity provided for $\lambda$ greater than 
or of order unity. On the other hand $g(\lambda)$ is another unknown function 
such that $g(\lambda) \rightarrow 0$ as $\lambda \rightarrow \infty$. 
There seems no reason not to assume that $g(\lambda)$ is of order unity 
when $\lambda$ is of order unity. 

Consider again varying the coupling constant down from infinity. The 
first term in \eqref{dilfflex} dominates the evolution of the temperature 
when the coupling is large enough (i.e for $g(\lambda) \ll T^{-1} \partial_0 \phi$ )
and \eqref{eqt} applies in this domain. On the other hand once $g(\lambda)$
increase above $T^{-1}\partial_0 \phi$, the second term in \eqref{dilfflex} 
dominates the evolution of the temperature. The details of the 
subsequent evolution depend on the precise form of the unknown function 
$g(\lambda)$. 

In order to get a sense for how this works, however let us assume that 
$g(\lambda)=K/\lambda^a$ for some positive power $a$ (this would be the expectation 
for $g(\lambda)$ at large $\lambda$ if it was generated at finite order in the $\alpha'$ 
expansion). In this case the second term of \eqref{dilfflex} dominates over the first 
when $\lambda \leq \lambda_0 \sim (\frac{T}{{\dot \phi}})^{1/a}$. For lower values of 
$\lambda$ we can ignore the first term in \eqref{dilfflex}, and then that equation is easily 
integrated. We find that, in lowering $\lambda$ from $\lambda_0$ to $\lambda_{min}$, 
the temperature changes by a factor $e^{-\frac{K}{a} (\frac{1}{\lambda_{min}^a}-\frac{1}{\lambda_0^a})}$. 
Now $\frac{\dot \phi}{T}$ must be small in order for fluid dynamics to apply. It follows that 
$\lambda_0 \gg \lambda_{min}$ when $\lambda_{min}$ is of unit order, so that lowering $\lambda$ from $\lambda_0$ to $\lambda_{min}$ changes the temperature of the fluid 
by a finite factor that is is independent of $\lambda_0$. 

In summary, if the coupling is lowered very slowly from $\lambda$ to $\lambda_0$ 
the temperature is almost unchanged. Subsequent lowering of the coupling from 
$\lambda_0$ to $\lambda_{min}$ changes the temperature by a fixed finite factor that depends 
on $\lambda_{min}$ but not on $\lambda_0$ or other details of the process. We conclude that 
the principal conclusions of the previous subsection - namely that the dilaton may slowly 
be lowered from an arbitrarily large though fixed value to a value of 
order one with a finite (though, in this case, not arbitrarily small) change temperature - 
carry through unchanged. This appears to be an interesting statement about the dual bulk 
evolution. We have argued that nothing particularly dramatic happens to a 
spacetime  - at least if it is at large finite temperature - as it 
traverses through a region of string scale curvature. We emphasize that it was very important 
to our analysis that we were at high temperature; the fluid dynamical analysis presented 
here sheds no light on the evolution of the same bulk geometries at zero temperature.

In fact one can try and go even further. After having reduced the dilaton to a value where,
$g_s N\sim O(1)$,  we can reduce it even more till $g_s N \ll 1$. 
At this stage the notion of spacetime geometry has completely broken down and the correct description
is in terms of a perturbative Yang Mills theory. Thereafter one can increase the dilaton,  till in the 
far future it meets the condition,  $g_s N\gg 1$.  As long as the driving force due to the dilaton is 
slowly enough varying, one expects that the boundary theory will be well defined, and the initial 
state we start with can be continued past the highly curved region to match with a smooth 
geometry in the future. 

Now  fluid mechanics can continue to be a good description, at small $g_sN$, provided, 
the function $h(\lambda)$ does not blow up   \footnote{
This seems very reasonable and can be checked against a perturbative calculation.} as $\lambda \rightarrow 0$,
 and, the temperature varies slowly enough, satisfying the condition, 
\be
\label{valf2}
l_{mfp} \frac{\dot{T}}{T} \ll 1,
\ee
(where $l_{mfp}$ stands for the mean free path).
The condition
\eqref{valf2} is expected to translate into
\be
\label{weakval}
\frac{\dot{T}}{T^2} \ll \lambda
\ee
at small $\lambda$, a  stronger condition than \eqref{fvalid} which holds
at $\lambda$ of unit order or greater. (This expectation follows
from the estimate,  that in perturbative Yang Mills theory
\be
\label{lmf3}
l_{mfp}\sim \frac{1}{(g_sN)^2 T}.
\ee
as opposed to $l_{mfp} \sim \frac{1}{T}$ at strong coupling).
Consequently, requiring that fluid mechanics is a good effective theory in the 
perturbative regime puts more stringent limits on how 
fast the dilaton can be varied. From eq.(\ref{valf2}), eq.(\ref{lmf3}), we  get, 
\be
\label{cond4}
\frac{d( \frac{1}{e^\phi N})}{dt} \ll T.
\ee
Nonetheless, using a dilaton profile  which varies slowly enough to  meet this 
more stringent condition,  we can construct solutions where two asymptotically smooth regions
of spacetime are connected by an intermediate region which is highly curved (dual to 
Yang Mills theory at any specified - though nonzero - coupling). 
This intermediate region  admits no conventional description in terms of spacetime, but we 
expect it to be  well described by fluid mechanics 
\footnote{Also it is worth noting that once the dilaton is small enough,  we 
could more directly analyze the 
time evolution in perturbation theory, even when fluid mechanics is not valid. This may even 
allow us to study evolutions that continue to strictly zero field theory 
coupling in finite time.}.

As must be clear to the reader, the discussion this subsection makes several assumptions and is tentative in some respects. In particular we reiterate that it makes assumptions that are  difficult
to directly verify: in particular the assumption that there are no phase transitions in high temperature 
${\cal N}=4$ theory as $g_sN$ is varied from large values to $O(1)$, or smaller values.

\subsection{Spatially varying Dilaton}\label{subs:spacevardil}

Finally we consider a spatially varying dilaton as a forcing function.
For simplicity we take the dilaton to depend on only one spatial coordinate, $x$.
We also set, the $u^y, u^z,$ components of the four-velocity to vanish.  
As in our discussion of the time dependent case above, we start with the perfect-fluid stress tensor, 
eq.(\ref{pfs}), and leading dilaton forcing function. Eq.(\ref{dilffl}) then gives rise to two condition.
From, the $\nu=0$ component we get,
\be
\label{nuzeroc}
\frac{d(T^4 u^0u^x)}{dx}=0.
\ee
And, from the $\nu=x$ component, 
\be
\label{nux}
\frac{d(T^4(4(u^x)^2 +1)}{dx}=-\frac{T^3}{\pi }u^x (\phi')^2,
\ee
where prime indicates derivative with respect to $x$. 
In addition the condition, 
\be
\label{condc}
(u^0)^2-(u^x)^2=1,
\ee
is also valid. This yields three equations in the three variables, $T, u^0, u^x$. 
Eliminating two variables from eq.(\ref{nuzeroc}), eq.(\ref{condc}), and substituting in 
eq.(\ref{nux}) reduces the problem to quadrature.  
   
Here let us consider the case where where the spatial velocity is small, i.e.,
\be
\label{spvel}
u^x \ll 1,
\ee
and work, to begin with, to linear order in $u^x$. 
To this order, $u^0=1$, and from eq.(\ref{nuzeroc}),
\be
\label{valux}
u^x=\frac{c}{T^4},
\ee
where $c$ is a constant. 
 Eq,(\ref{nux}), now gives,
\be
\label{soltspa}
\frac{T^5(x)}{T_i^5}= 1-\frac{5}{4 \pi}\frac{u_i^x}{T_i} \int_{-\infty}^x (\phi')^2 dx.
\ee
where $T_i, u_i^x$ stand for the temperature and velocity at $x\rightarrow -\infty$.
Without loss of generality we can choose conventions so that $u_i^x>0$. Then we see that 
as $x$ increases,  the temperature decreases and the velocity increases. From, eq.(\ref{valux}), 
eq.(\ref{soltspa}), we see that the length scale of
 variation of the temperature
and velocity,  are set by the forcing function of the dilaton. We denote this length scale by $L$ below.  

The analysis can be improved by working to quadratic order in $u^x$. In general also the dilaton 
will depend on all spatial coordinates, this will lead to spatially varying velocities and to 
viscosity effects being important. 
 The Reynolds number  for a fluid flow is given by,
\be
\label{ren}
R_e\sim \Delta u^x TL,
\ee
where $\Delta u^x$ is the variation in the velocity along the flow. 
From, eq.(\ref{valux}), eq.(\ref{soltspa}), we see that, for $\Delta \phi \sim O(1)$, 
$\Delta u^x \sim u^x_i \frac{\Delta T}{T} \sim (u^x_i)^2 \frac{1}{TL}$.  
This gives Reynolds numbers of order unity or smaller. Again this is in line with the 
expectations spelt out in the introduction. The smallness of the Reynolds number of this 
flow is connected to the fact that the dilaton yields a forcing function that is of 
second rather than first order in derivatives.

Corrections to these flows can be systematically computed, using the corrections to the stress tensor,
eq.(\ref{stetc}), and forcing function, eq.(\ref{enmomforce}).

\section{Discussion}

In this paper we have used the AdS/CFT correspondence to determine
the metric dual to an arbitrary flow of the fluid of a conformal field 
theory on an arbitrary weakly curved four manifold, with an arbitrary 
slowly varying coupling constant. We have also explicitly constructed
the event horizon of these metrics in the derivative expansion, and used 
gravitational ideas to propose the construction of a dual entropy current
whose local increase is guaranteed by the area increase theorem for 
black holes in general relativity. 

We have tested the constructions of this paper against an exact 
class of asymptotically globally AdS$_5$ solutions of Einstein's equations, 
namely rotating black holes in global AdS$_5$. We have demonstrated that 
these solutions admit a remarkably simple rewriting in fluid dynamical 
terms, and have verified that they agree in detail with the general 
constructions of our paper when expanded out to second order in a derivative 
expansion. 

The connection of large rotating black holes to fluid dynamics was previously
partially explored in \cite{Bhattacharyya:2007vs}. The authors of that 
paper noted that the stress tensor dual to rotating black hole solutions 
was exactly reproduced, to appropriate order in the derivative expansion, 
by the predictions of fluid dynamics. However these authors also noted 
an apparent first order discrepancy between the properties of charged 
rotating black holes\cite{Chong:2005da,Cvetic:2004ny,Chong:2005hr,Chong:2006zx}
 and the predictions of charge fluid dynamics. 
A generalization of the calculations of this paper to account for 
fluid charge would allow us to pinpoint the source of this worrying 
discrepancy.

Turning to another issue, we believe  that it should 
be  possible, within the approximations of this paper, 
to choose a time dependent fluctuations of a flat space metric to stir 
the conformal fluids studied of this paper into steady motions with 
high Reynolds numbers. Even using the forcing functions provided by 
linearized fluctuations of the metric away from flat space (see the 
introduction) we believe it should be possible, for instance, to stir a 
fluid on $R^3$ or the sphere into configurations that resemble classic 
experiments that display turbulence\footnote{A time dependent $h_{0i}$ 
provides a forcing function in the $i^{th}$ direction.}. 
Consequently, the map from fluid dynamics to metrics in this paper should 
yield the spacetime dual to a turbulent fluid flow. The understanding of 
apparently universal exponents in turbulent motions remains one of the 
outstanding unsolved problems of statistical physics.  It is conceivable 
that the dual spacetime perspective could permit qualitatively new 
insights for such flows. 

Of course transient turbulence occurs in fluid flows even without forcing. For instance, 
a configuration of our conformal fluid at temperature $T$ on an $S^3$ of radius $R$, 
and with (varying) velocities of unit order has Reynolds number of order $TR$ and 
so presumably undergoes turbulent motion. Now all such fluid configurations eventually 
settle down into (presumably unique) non dissipative solutions of of fluid 
dynamics with given conserved energies and angular momenta; the rigid rotations 
studied in \cite{Bhattacharyya:2007vs} and in section \ref{adskerr:sec} of this paper. Let us translate
this expectation to bulk language. Consider an arbitrary gravitational configuration, 
in global AdS$_5$, with appropriately large energy and angular momenta. 
Such a configuration will undergo gravitational collapse and 
 eventually settle 
down into one of the rotating black holes studied in  \cite{Bhattacharyya:2007vs}. 
The gravitational dynamics of this settling down process could be dual to 
a turbulent fluid flow. It may be interesting to pursue this connection further. 
 
Finally, we end this paper with an amusing thought. Recall that we have been able to 
construct a bulk dual to fluid flow in an arbitrary boundary metric. 
Now the boundary metric can, in particular, be chosen to have a horizon; 
for example it could be taken to be the metric of a 4 dimensional 
Schwarzschild black hole(See \cite{Chamblin:1999by} for an example). It may be interesting to investigate the 
bulk dual description of the boundary fluid falling through horizon 
of this metric.

\acknowledgments

{\bf Acknowledgments}

We would like to acknowledge useful discussions and correspondences
with J.Bhattacharya, S. Das, V. Hubeny, G.Mandal, M. Rangamani and M.Van Raamsdonk. We would also like to acknowledge useful discussion with the students in the TIFR theory room. Several of the calculations of this paper were performed using the excellent Mathematica package \texttt{diffgeo.m} developed by Matthew Headrick. The work of S.M. and S. P. T. was supported in part by Swarnajayanti Fellowships. The work of S.R.W was supported in part by a J. C. Bose Fellowship. We would all also like to acknowledge our debt to the people of India for their generous and steady support to research in the basic sciences.

\appendix
\section{Force on the Boundary fluid due to a Varying Dilaton}\label{app:forcing}
In this appendix, we calculate the force that a varying dilaton field applies on a fluid at the boundary. To that end, we start with the bulk expression for the boundary energy-momentum tensor
\begin{equation}\label{Tmunu:eq}
\begin{split}
16\pi G_5 T^A_B &= \lim_{r\rightarrow\infty} r^4\left(2(K_{CD}h^{CD} h^{A}_{B} - K^A_B)\right.\\
&\left.+\bar{\mathcal{G}}^A_B-6 h^{A}_{B}  -\frac{1}{2}\left(\bar{\nabla}^A\Phi\bar{\nabla}_B\Phi-\frac{h^A_B h^{CD}}{2} \bar{\nabla}_C \Phi \bar{\nabla}_D \Phi \right) \right)
\end{split}
\end{equation}
where $h_{AB}, K_{AB}$ and $\bar{\mathcal{G}}_{AB}$ are respectively the induced metric, the extrinsic curvature and the Einstein tensor of the constant r hypersurface. $\bar{\nabla}$ is the covariant derivative corresponding to the constant r hypersurface in the bulk.

To compute the divergence $f_B= \bar{\nabla}_A T^{A}_B$, we proceed as follows -  from Gauss-Codazzi-Mainardi relations (See, for example, Eqn.10.2.24 of \cite{1984ucp..book.....W}), we have
\begin{equation}\label{codacci:eq}
2\bar{\nabla}_A\left(K_{CD}h^{CD} h^{A}_{B} - K^A_B\right)=-2 R_{CD}g^C_B n^D = -\nabla_C\Phi \nabla_D\Phi g^C_B n^D = -\bar{\nabla}_B\Phi \nabla_n\Phi  
\end{equation}
where in the last step we have used the Einstein equations in the bulk. Further, we also have
\begin{equation}\label{gradphi:eq}
-\frac{1}{2}\bar{\nabla}_A\left(\bar{\nabla}^A\Phi\bar{\nabla}_B\Phi-\frac{h^A_B h^{CD}}{2} \bar{\nabla}_C \Phi \bar{\nabla}_D \Phi \right)= -\frac{1}{2}\bar{\nabla}^2\Phi \bar{\nabla}_B\Phi
\end{equation}

Using these two equations along with the reduced Bianchi identity $\bar{\nabla}_A\bar{\mathcal{G}}^{AB}=0$, we get
\begin{equation}
16\pi G_5 \bar{\nabla}_A T^{A}_B = -\bar{\nabla}_B\phi\left(\nabla_n\Phi+\frac{1}{2}\bar{\nabla}^2\Phi\right)
\end{equation}
Multiplying by $r^4$ and using $16\pi G_5 e^{-\phi} \mathcal{L} = - \lim_{r\rightarrow\infty} r^4\left(\nabla_n\Phi+\frac{1}{2}\bar{\nabla}^2\Phi\right)$ we get the following relation among the boundary variables
\begin{equation}
\nabla_\mu T^\mu_\nu = e^{-\phi} \mathcal{L} \nabla_\nu\phi
\end{equation}

\section{AdS Kerr}\label{app:adskerr}
In order to rewrite AdS Kerr metric in the fluid dynamical form, we find 
it convenient to re-express it as
\begin{equation}\label{ads5kerralt1:eq}
\begin{split}
ds^2 &= 2dr\left( dt-\frac{\omega_1 \sin^2\Theta}{1-\omega_1^2}d\phi 
-\frac{\omega_2 \cos^2\Theta}{1-\omega_2^2}d\psi \right)-(r^2 + 1)\frac{\Delta_\Theta dt^2}{(1-\omega_1^2)(1-\omega_2^2)} \\
&+ \rho^2 \frac{d\Theta^2}{\Delta_\Theta} 
+ (r^2 + \omega_1^2)\frac{\sin^2\Theta}{1-\omega_1^2}(d\phi+\omega_1 dt)^2 
+ (r^2 + \omega_2^2)\frac{\cos^2\Theta}{1-\omega_2^2}(d\psi+\omega_2 dt)^2 \\
&+ \frac{2 m}{\rho^2}\left( dt-\frac{\omega_1 \sin^2\Theta}{1-\omega_1^2}d\phi -\frac{\omega_2 \cos^2\Theta}{1-\omega_2^2}d\psi \right)^2 \\
&= 2dr\left( dt-\frac{\omega_1 \sin^2\Theta}{1-\omega_1^2}d\phi 
-\frac{\omega_2 \cos^2\Theta}{1-\omega_2^2}d\psi \right)\\
&+r^2\left(-\frac{\Delta_\Theta dt^2}{(1-\omega_1^2)(1-\omega_2^2)}
+ \frac{d\Theta^2}{\Delta_\Theta} 
+ \frac{\sin^2\Theta}{1-\omega_1^2}(d\phi+\omega_1 dt)^2 
+ \frac{\cos^2\Theta}{1-\omega_2^2}(d\psi+\omega_2 dt)^2 \right) \\
&-\frac{\Delta_\Theta dt^2}{(1-\omega_1^2)(1-\omega_2^2)} + (\omega_1^2 \cos^2\Theta +\omega_2^2 \sin^2\Theta) \frac{d\Theta^2}{\Delta_\Theta} \\
&+ \frac{\omega_1^2\sin^2\Theta}{1-\omega_1^2}(d\phi+\omega_1 dt)^2 
+ \frac{\omega_2^2\cos^2\Theta}{1-\omega_2^2}(d\psi+\omega_2 dt)^2 \\
&+ \frac{2 m}{\rho^2}\left( dt-\frac{\omega_1 \sin^2\Theta}{1-\omega_1^2}d\phi -\frac{\omega_2 \cos^2\Theta}{1-\omega_2^2}d\psi \right)^2\\
\end{split}
\end{equation}
This metric may be rewritten as 
\begin{equation}\label{ads5kerralt2:eq}
\begin{split}
ds^2 &= -2 u_\mu dx^\mu dr + r^2 g_{\mu\nu} dx^\mu dx^\nu  + \frac{2 m}{\rho^2} u_\mu u_\nu dx^\mu dx^\nu +\Sigma_{\mu\nu}dx^\mu dx^\nu
\end{split}
\end{equation}
where  
\begin{equation}\label{Sigma:eq}
\begin{split}
\Sigma_{\mu\nu}dx^\mu dx^\nu &\equiv-\frac{\Delta_\Theta dt^2}{(1-\omega_1^2)(1-\omega_2^2)} + (\omega_1^2 \cos^2\Theta +\omega_2^2 \sin^2\Theta) \frac{d\Theta^2}{\Delta_\Theta}\\
&+ \frac{\omega_1^2\sin^2\Theta}{1-\omega_1^2}(d\phi+\omega_1 dt)^2 
+ \frac{\omega_2^2\cos^2\Theta}{1-\omega_2^2}(d\psi+\omega_2 dt)^2 
\end{split}
\end{equation}
The first three terms in \eqref{ads5kerralt2:eq} are simply the ansatz
\eqref{genbbn2} while the last term, proportional to $\Sigma_{\mu\nu}$ 
represents the derivative corrections \eqref{pertform} to that ansatz.  

It is possible to express $\Sigma_{\mu\nu}$ entirely in terms of 
Weyl-covariant curvatures and Weyl covariant derivatives of the velocity. 

The non-zero Weyl covariants associated with this fluid configuration are 
\begin{equation}\label{weylREU:eq}
\begin{split}
\mathcal{A}_\mu &= 0\ ;\qquad\ \sigma_{\mu\nu}=0\ ;\qquad\ C_{\mu\nu\lambda\sigma} = 0 \\
\mathcal{R}&=6\left(1+\omega_1^2+\omega_2^2-3(\omega_1^2 \cos^2\Theta +\omega_2^2 \sin^2\Theta)\right)\\ \omega_{\mu\nu}\omega^{\mu\nu} &= 2 (\omega_1^2 \cos^2\Theta +\omega_2^2 \sin^2\Theta) \\
\frac{1}{2} \omega_{\mu\nu} dx^\mu \wedge dx^\nu &= \sin\Theta\cos\Theta d\Theta\wedge\left( \frac{\omega_1}{1-\omega_1^2} d\phi
-\frac{\omega_2}{1-\omega_2^2} d\psi\right)\\
\mathcal{D}_\lambda \omega^\lambda{}_\mu dx^\mu &= P^\nu_\mu\mathcal{D}_\lambda \omega^\lambda{}_\nu dx^\mu + \omega_{\alpha\beta}\omega^{\alpha\beta} u_\mu dx^\mu \\
&= -2dt(\omega_1^2 \cos^2\Theta +\omega_2^2 \sin^2\Theta)\\
&-2\Delta_\Theta\left( \frac{\omega_1\sin^2\Theta}{1-\omega_1^2}d\phi+
\frac{\omega_2\cos^2\Theta}{1-\omega_2^2}d\psi\right)\\
& +2 (\omega_1^2 - \omega_2^2) \sin^2\Theta\cos^2\Theta \left(
\frac{\omega_1 d\phi}{1 - \omega_1^2}- 
\frac{\omega_2 d\psi}{1 - \omega_2^2}\right)\\
P^\nu_\mu\mathcal{D}_\lambda \omega^\lambda{}_\nu dx^\mu &= -2\left( \frac{\omega_1\sin^2\Theta}{1-\omega_1^2}d\phi+
\frac{\omega_2\cos^2\Theta}{1-\omega_2^2}d\psi \right)\\
& +2 (\omega_1^2 - \omega_2^2) \sin^2\Theta\cos^2\Theta \left(
\frac{\omega_1 d\phi}{1 - \omega_1^2}- 
\frac{\omega_2 d\psi}{1 - \omega_2^2}\right)\\
P_{\mu\nu} dx^\mu dx^\nu &= \frac{d\Theta^2}{\Delta_\Theta}+ \frac{\sin^2\Theta}{1-\omega_1^2}d\phi^2+
\frac{\cos^2\Theta}{1-\omega_2^2}d\psi^2 \\
&-\sin^2\Theta\cos^2\Theta \left(
\frac{\omega_1 d\phi}{1 - \omega_1^2}- 
\frac{\omega_2 d\psi}{1 - \omega_2^2}\right)^2 \\
\omega_\mu{}^\lambda\omega_{\lambda\nu} dx^\mu dx^\nu &= -(\omega_1^2 \cos^2\Theta +\omega_2^2 \sin^2\Theta)\frac{d\Theta^2}{\Delta_\Theta}\\
&- \Delta_\Theta \sin^2\Theta\cos^2\Theta \left(
\frac{\omega_1 d\phi}{1 - \omega_1^2}- 
\frac{\omega_2 d\psi}{1 - \omega_2^2}\right)^2\\
\end{split}
\end{equation}
\begin{equation*}
\begin{split}
\left(\omega_\mu{}^\lambda\omega_{\lambda\nu} +\frac{\omega_{\alpha\beta}\omega^{\alpha\beta}}{3} P_{\mu\nu} \right)& dx^\mu dx^\nu = -\frac{1}{3}(\omega_1^2 \cos^2\Theta +\omega_2^2 \sin^2\Theta)\frac{d\Theta^2}{\Delta_\Theta}\\
&+\frac{2}{3}(\omega_1^2 \cos^2\Theta +\omega_2^2 \sin^2\Theta)\left( \frac{\sin^2\Theta}{1-\omega_1^2}d\phi^2+
\frac{\cos^2\Theta}{1-\omega_2^2}d\psi^2 \right)\\
&+\frac{\Delta_\Theta+2}{3}\sin^2\Theta\cos^2\Theta \left(
\frac{\omega_1 d\phi}{1 - \omega_1^2}- 
\frac{\omega_2 d\psi}{1 - \omega_2^2}\right)^2 \\
\end{split}
\end{equation*}

Using the above expressions, one can rewrite $\Sigma_{\mu\nu}$ completely 
in terms of the Weyl-covariant derivatives of the fluid velocity 
and Weyl covariant boundary curvatures: we find 
\begin{equation}\label{SigmaWeyl:eq}
\begin{split}
\Sigma_{\mu\nu}dx^\mu dx^\nu &= -\left(u_\mu \mathcal{D}_\lambda\omega^\lambda{}_\nu + \omega_\mu{}^\lambda\omega_{\lambda\nu} + \frac{\mathcal{R}}{6} u_\mu u_\nu \right)dx^\mu dx^\nu \\
\text{and} &\qquad\ \rho^2 = r^2 + \frac{1}{2} \omega_{\alpha\beta}\omega^{\alpha\beta}= r^2\left(1+\frac{1}{2r^2} \omega_{\alpha\beta}\omega^{\alpha\beta}\right)
\end{split}
\end{equation}

\section{Notation} \label{app:notation}
We work in the mostly positive, $(-++\ldots)$, signature.  The dimensions of the spacetime in which the conformal fluid lives is denoted by $d$. In the context of AdS/CFT, the dual AdS$_{d+1}$ space has $d+1$ spacetime dimensions.  

Latin alphabets $A,B,\ldots$ are used to denote the $d+1$ dimensional bulk indices which range over $\{r,0,1,\ldots,d-1\}$. Lower Greek letters $\mu,\nu,\ldots$ indices range over $\{0,1,\ldots,d-1\}$ .   The co-ordinates in the bulk are denoted by $X^A$ which is often split into a radial co-ordinate $r$ and $x^\mu$. The dilaton field in the bulk is denoted by $\Phi$ and its value in the boundary is denoted by $\phi$.  

Our convention for the Riemann curvature tensor is fixed by the relation
\begin{equation}
[\nabla_\mu,\nabla_\nu]V^\lambda=-R_{\mu\nu\sigma}{}^{\lambda}V^\sigma . 
\end{equation}

In Table \ref{nottable}, we list the physical meaning and the definitions of various quantities used in the text.


\TABLE{\label{nottable}
\caption{Conventions used in the text,with reference to the equations defining them where appropriate.}
    \begin{tabular}{||c|l||c|l||}
    \hline
    {\bf Symbol} & {\bf Definition} & {\bf Symbol} & {\bf Definition} \\
    \hline
    $d$ & dimensions of boundary & $G_{AB}$ & Bulk metric\\
    & & $g_{\mu\nu}$ & Boundary metric  \\
    \hline \hline
%
    $T$ & Fluid temperature & $\eta$ & Shear viscosity  \\
    $T^{\mu\nu}$ & Energy-momentum tensor & $J^\mu_S$ & Entropy current \\
    $u^\mu$ & Fluid velocity ($u^\mu u_\mu =-1$) &  $P^{\mu\nu}$ & Projection tensor, $
    g^{\mu\nu}+u^\mu u^\nu$ \\
    $a^\mu$ & Fluid acceleration,   &$\vartheta$&  Fluid expansion, \\
    $\sigma_{\mu\nu}$ & Shear strain rate,  & $\omega_{\mu\nu}$ & Fluid vorticity,  \\
    \hline \hline
%
     $\mathcal{D}_\mu$ & Weyl-covariant derivative  & $\mathcal{A}_\mu$ & See  \eqref{weylmet:eq}\\
     $R_{\mu\nu\lambda}{}^{\sigma}$ & Riemann tensor &      $\mathcal{F}_{\mu\nu}$ & $\nabla_\mu\mathcal{A}_\nu-\nabla_\nu\mathcal{A}_\mu$ \\
     $C_{\mu\nu\lambda\sigma}$ & Weyl curvature & &\\
     \hline
\end{tabular}
}

\bibliographystyle{JHEP}
\bibliography{force}

\end{document}